\documentclass[aps,pra,reprint,showpacs,preprintnumbers,amsmath,amssymb]{revtex4-1} 
\usepackage{amsmath}
\usepackage{graphicx}
\usepackage{amssymb}
\usepackage{textcomp}
\usepackage{color}
\usepackage{xcolor}
\usepackage{float}
\usepackage{natbib}
\usepackage{dcolumn}
\usepackage{ragged2e}
\usepackage{color,soul}
\usepackage{array}
\usepackage{dcolumn}
\usepackage{url}
\usepackage[colorlinks=true,linkcolor=black,citecolor=black,urlcolor=blue,bookmarks=false,hypertexnames=false]{hyperref}
\newcolumntype{.}{D{.}{.}{-1}}
\newcolumntype{L}[1]{>{\raggedright\let\newline\\\arraybackslash\hspace{0pt}}m{#1}}
\newcolumntype{C}[1]{>{\centering\let\newline\\\arraybackslash\hspace{0pt}}m{#1}}
\newcolumntype{R}[1]{>{\raggedleft\let\newline\\\arraybackslash\hspace{0pt}}m{#1}}

\begin{document}

\title{Interaction potentials and ultracold scattering cross sections for the $^7$Li$^+$-$^7$Li ion-atom system}
\author{A. Pandey}
\email{amrendra.pandey@universite-paris-saclay.fr} 
\affiliation{Raman Research Institute, Light and Matter Physics, Sadashivanagar, Bangalore 560080, India}
\author{M. Niranjan}
\affiliation{Raman Research Institute, Light and Matter Physics, Sadashivanagar, Bangalore 560080, India}
\author{N. Joshi}
\affiliation{Raman Research Institute, Light and Matter Physics, Sadashivanagar, Bangalore 560080, India}
\author{S. A. Rangwala}
\affiliation{Raman Research Institute, Light and Matter Physics, Sadashivanagar, Bangalore 560080, India}

\author{R. Vexiau}
\affiliation{Universit\'e Paris-Saclay, CNRS, Laboratoire Aim$\acute{\text e}$ Cotton, Orsay, 91400, France}
\author{O. Dulieu}
\affiliation{Universit\'e Paris-Saclay, CNRS, Laboratoire Aim$\acute{\text e}$ Cotton, Orsay, 91400, France}

\date{\today}

\begin{abstract} 
We calculate the isotope independent Li$^+$-Li potential energy curves for the electronic ground and first excited states. Scattering phase shifts and total scattering cross section for the $^7$Li$^+$-$^7$Li collision are calculated with emphasis on the ultra-low energy domain down to the $s$-wave regime. 
The effect of physically motivated alterations on the calculated potential energy curves is used to determine the bound of accuracy of the low-energy scattering parameters for the ion-atom system. It is found that the scattering length for the A$^2\Sigma_u^+$ state, $a_u$ = 1325 a$_0$, is positive and has well-constrained bounds. For the X$^2\Sigma_g^+$ state, the scattering length, $a_g$ = 20465 a$_0$ has a large magnitude as it is sensitive to the restrained change of the potential, due to the presence of a vibrational state in the vicinity of the dissociation limit. 
\end{abstract}
\maketitle

\section{Introduction}\label{sec:one}

Experimental research on ion-atom interactions in dilute, trapped gas systems at ultracold temperatures is rapidly evolving towards detailed probes of the quantum dynamics of the resulting products \cite{grier2009observation,zipkes2010trapped,schmid2010dynamics,meir2016dynamics,ratschbacher2012controlling,haze2013observation,haze2015charge,ravi2012cooling,sivarajah2012evidence,dutta2017collisional,hall2011light,schowalter2016blue,harter2012single,morita2018spin,sikorsky2018spin,wang2019controlling}. One of the main goals is to thermalize an atomic ion within the ultracold atomic gas \cite{zipkes2010trapped,schmid2010dynamics,ratschbacher2012controlling}. An atom and an ion mutually interact at large internuclear distance, R, through an attractive charge-induced-dipole potential behaving as $\sim -\alpha_d/(2\text{R}^4)$, where $\alpha_d$ is the static dipole polarizability of the neutral atom.  For energy E$\geq k_\text{B} \times 1$ mK, an ion-atom collision involves many partial waves $\ell$, due to the strongly attractive long-range nature of their interaction \cite{massey1972atom}, allowing a semiclassical description of the collision. Despite continuous progress regarding the precise control of the trapped ion motion, reaching the ultra-low relative energy regime (E/$k_\text{B}$ $\approx 1 \mu$K or lower) for ion-atom collisions is still challenging experimentally \cite{cetina2012micromotion,meir2016dynamics,haze2013observation,feldker2020buffer,ben2019strong}. At these energies, quantum effects emerge as only few partial waves contribute to the collision. Due to ion heating as a result of interactions and trap imperfections in  dynamical trapping, it is experimentally advantageous to investigate the full quantum regime at the highest possible temperatures \cite{meir2016dynamics,cetina2012micromotion}.

The lowest possible centrifugal barrier is induced by the $p$-wave ($\ell=1$) and has a height equal to $1/( 2\mu^2 \alpha_d)$ (in atomic units of energy). The $p$-wave barrier will be the highest for low reduced mass, $\mu$, thus opening the possibility to probe it at relatively high collision energy. For this reason, lithium is implemented in several ongoing experiments  \cite{harter2014cold,haze2013observation,haze2015charge,joger2017observation}. Most hybrid ion-atom trapping experiments use an alkaline-earth ionic species suitable for laser cooling, which aids the achievement of low ion-atom collision energies. The choice of a heteronuclear ion-atom combination, however, excludes the resonant charge exchange (RCE) mechanism, where an electron of the atom can be transferred to the ion without any energy release \cite{cote2000classical,cote2000ultracold,grier2009observation,lee2013measurement,ravi2012cooling,sivarajah2012evidence,dutta2018cooling}. In our previous experiments \cite{ravi2012cooling,dutta2018cooling}, we have consistently exploited the RCE in the study of ion-atom collisions. We therefore focus this study on the scattering properties of $^7$Li$^+$-$^7$Li as this is a light system, with isotopic abundance, for going toward the quantum regime, with  a $p$-wave barrier height of k$_B \times 2.98 \times 10^{-5}$ K.

In this paper, we perform calculations of the $^7$Li$^+$-$^7$Li interaction for the colliding ion and the atom when they are in their ground state. Specifically, we compute the \textit{ab initio}  potential energy curves (PECs) of the X$^2\Sigma_g^+$, the electronic ground state, and the A$^2\Sigma_u^+$, the first electronic excited state, of the Li$_2^+$ molecular ion using the multireference configuration interaction (MRCI) method and the best available basis sets. This is required despite the availability of previous high quality calculations, since there is a significant discrepancy ($\approx$ factor of 2) between the calculations for $a_g$ \cite{zhang2009near,schmid2018rydberg}, the scattering length for the X$^2\Sigma_g^+$ state, which determines the low-energy ion-atom scattering cross section. These molecular ion \textit{ab initio} PECs are smoothly matched to their physical asymptotic forms in the large-R range. We then derive the phase-shifts characterizing the $^7$Li$^+$-$^7$Li collision as functions of the energy. The resulting scattering lengths $a_g$ and $a_u$ of the X$^2\Sigma_g^+$ and A$^2\Sigma_u^+$ states, respectively, are both computed to be positive with $a_g \gg a_u$. Our results are consistent with previous studies on the X$^2\Sigma_g^+$ and A$^2\Sigma_u^+$ PECs \cite{schmidt1985ground,magnier1999potential,bouzouita2006ab,jasik2007calculation,zhang2009near,musial2015potential,rabli2017revised,nasiri2017benchmark}. A convergence criterion is developed to bound the range of uncertainty within which the values of scattering lengths, $a_g$ and $a_u$, are constrained. 
The total cross sections by computing the phase shifts are evaluated in section III. We finally provide recommended values for the cross sections and their bounds for the  $^7$Li$^+$-$^7$Li  system.

\section{L{i}$_2^+$ Potential energy curves}
\label{sec:PEC}
\subsection{\textit{Ab initio} Born-Oppenheimer Potentials}  

We compute the X$^2\Sigma_g^+$ and A$^2\Sigma_u^+$ states of Li$_2^+$ under the Born-Oppenheimer approximation using the MOLPRO package \cite{MOLPRO-WIREs}. The complete active space self-consistent field (CASSCF) and multireference configuration interaction (MRCI) with single and double excitations (-SD) methods are used. Full-valence type CASSCF wave functions, which consider all five electrons of Li$_2^+$ as active, are calculated and used as the reference functions for the MRCI calculations \cite{werner1988efficient}. We choose this approach as it is variational for the truncated configuration interaction (CI) expansion, to ensure convergence towards the true energies for both states with the basis-set size. The reference calculations are performed with the largest available basis set, namely augmented Dunning correlation-consistent, polarized valence, 5-zeta basis set, aug-cc-pCV5Z \cite{prascher2011gaussian}. 

Due to the large discrepancies between $a_g$ values reported in the literature \cite{zhang2009near,schmid2018rydberg}, we determine bounds for $a_g$ so that more precise calculations in the future should not supersede the conclusions drawn here. We first compute the atomic energies of Li and Li$^+$ in their ground state (Table \ref{tab:tab1Li+Li}). Various sizes of the aug-cc-pCVXZ basis sets are considered with X $\equiv$ D (double), T (triple), Q (quadruple), 5 (quintuple) referring to the largest excitation degree of the determinants. This allowed reaching a relative convergence better than 0.009\%. Our variational values are larger by 0.05\% in magnitude than the ones obtained in \cite{musial2015potential} using a Coupled-Cluster approach with single and double excitations (CCSD) and ANO-RCC+ basis set. As Li$^+$ and Li are small systems with two and three electrons, respectively, extremely precise atomic calculations can be performed. Our energies obtained with aug-cc-pCV5Z basis set differ from the best available variational calculations using Hylleraas coordinates \cite{drake2006springer,yan1995eigenvalues} by only 0.008\%, (Table \ref{tab:tab1Li+Li}), justifying that the choice of the aug-cc-pCV5Z set as an appropriate one for molecular calculations.

The sum of the electronic energies of Li$^+$ ($^1$S$_0$) and Li($^2$S$_{1/2}$), from Table I, and the energy of the dissociation limit obtained from the molecular calculation, $E_\infty$, Table \ref{tab:tab2Li2+basis}), exhibit a small difference ( 0.004 cm$^{-1}$) which is assigned to the basis set superposition error (BSSE). We calculated this effect for the Li atom, using effective core potential and core polarization potential with one valence electron (see the method labeled Th$_2$ further on). The BSSE amounts to less than 0.2 cm$^{-1}$ at the equilibrium distance, $R_e$, and to 0.006 cm$^{-1}$ at R = 50 a$_0$ (a$_0$ is the Bohr radius). Hence for the scattering calculation this correction is not incorporated into the potentials.

\begin{figure}[t]
\includegraphics[width=0.490\textwidth, angle = 0]{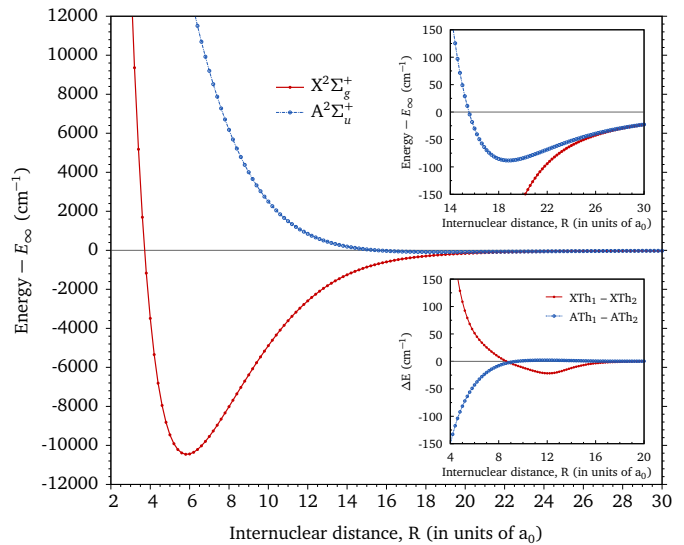}
\caption{$^7$Li$_2^{+}$ potential energy curves X$^2\Sigma_g^+$ and A$^2\Sigma_u^+$, computed in the present work (MRCI-SD with aug-cc-pCV5Z basis set), and respectively denoted as X${\text{Th}_1}$ and A${\text{Th}_1}$. The minima of the A$^2\Sigma_u^+$ curve is shown in the upper inset. The energy differences $\Delta$E with the curves calculated using the approach of \cite{aymar2005calculation} (denoted by X${\text{Th}_2}$ and A${\text{Th}_2}$), are shown in the lower inset. The corresponding spectroscopic constants are listed in Table \ref{tab:tab2Li2+basis}.}
\label{fig:PECLi2+a}
\end{figure}

\setlength{\tabcolsep}{3.5pt}
\begin{table}[t]
\vspace{2pt}
\begin{tabular}{lllc}  
\hline
Li$^+$ ($^1$S$_0$)& Li ($^2$S$_{1/2}$) & Li$^+$+Li  & Ref.             \\		\hline
-7.26922697       &-7.46607917  &  -14.73530614    & X $\equiv$ D \\
-7.27690629       &-7.47457432  &  -14.75148061   & X $\equiv$ T \\
-7.27870222       &-7.47670230  &  -14.75540452     & X $\equiv$ Q \\
-7.27933195       &-7.47740563  &  -14.75673758   & X $\equiv$ 5 \\
-7.275561           &-7.473553      &  -14.74911400  &\cite{musial2015potential}\\
-7.27991339$^a$    & -7.47806032310$^b$  & -14.7579737131  &\cite{drake2006springer}$^a$,\cite{yan1995eigenvalues}$^b$ \\ 
\hline
\end{tabular}
\caption{Total electronic energies (in a.u.) of the Li$^+$($^1$S$_0$) and Li($^2$S$_{1/2}$) ground states, their sum, Li$^+$+Li, obtained from the present MRCI-SD calculations with increasing size of basis sets from aug-cc-pCVXZ, with X $\equiv$ D, T, Q, 5. Another calculation using coupled cluster method, EA-EOM-CCSD with ANO-RCC+ basis sets \cite{musial2015potential} is provided for comparison. Two separate calculations on Li$^+$ and Li representing the non-relativistic variational calculations using Hylleraas coordinates are also reported \cite{drake2006springer,yan1995eigenvalues}.}
\label{tab:tab1Li+Li}
\end{table}

\setlength{\tabcolsep}{6.0pt}
\begin{table*}[t]
\vspace{2pt}
\begin{tabular}{ccccccccC{3.0cm}}
\hline
Electronic & $E_\infty$ & $\Delta E_\infty$ & $E_e$ & $\Delta E_e$ & $R_e$ & $D_e$ & R$_{\textrm{in}}$ & aug-cc-pCVXZ\\
State & (a.u.) &  (\%) & (a.u.) & (\%) & (units of a$_0$) & (cm$^{-1}$)& (units of a$_0$) &  basis sets, with X $\equiv$ \\
\hline
X$^2\Sigma_g^+$ & -14.73530934 &          & -14.78224306 &          &  5.940 & 10300.76 & 3.758 & D\\
                & -14.75148110 & 0.1097 & -14.79891577 & 0.1128 &  5.875 & 10410.70 & 3.723 & T\\
                & -14.75540520 & 0.0266 & -14.80300996 & 0.0277 &  5.865 & 10448.03 & 3.715 & Q\\
                & -14.75673756 & 0.0090 & -14.80438625 & 0.0093 &  5.858 & 10458.58 & 3.713 & 5\\

A$^2\Sigma_u^+$ & -14.73530934 &          & -14.73570944 &          & 18.939   & 87.81 & 15.630 & D\\
                & -14.75148110 & 0.1097 & -14.75188381 & 0.1098 & 18.839 & 88.38 & 15.563 & T\\
                & -14.75540520 & 0.0266 & -14.75580764 & 0.0266 & 18.818 & 88.32 & 15.545 & Q\\
                & -14.75673756 & 0.0090 & -14.75714022 & 0.0090 & 18.799 & 88.37 & 15.540 & 5\\
\hline
\end{tabular}
\caption{Dissociation limit, $E_\infty$, its convergence with basis sets, $\Delta E_\infty$, total energy $E_e$ at $R_e$, its convergence with basis sets, $\Delta E_e$, equilibrium distance, $R_e$, well depth, $D_e$, and repulsive wall position R$_{\textrm{in}}$ of the X$^2\Sigma_g^+$ and A$^2\Sigma_u^+$ PECs of $^7$Li$_2^+$. Results for various basis sets aug-cc-pCVXZ, with X $\equiv$ D, T, Q, 5 are displayed.} 
\label{tab:tab2Li2+basis} 
\end{table*}

In order to provide a convergence criterion on potential energies, we compute the \textit{ab initio} X$^2\Sigma_g^+$ and A$^2\Sigma_u^+$ PECs with a series of aug-cc-pCVXZ basis sets (with X $\equiv$ D, T, Q, 5) in the [2 a$_0$ - 50 a$_0$] internuclear distance range, with a 0.2 a$_0$ step. They correlate to the lowest asymptotic limit Li$^+$($^1$S$_0$) + Li($^2$S$_{1/2}$). We report in Table \ref{tab:tab2Li2+basis} the total potential energy  $E_\infty$ for R $\rightarrow \infty$, i.e. at the dissociation limit (see section \ref{sec:LR}), and $E_e$ at the equilibrium distance, $R_e$, the well depth $D_e$ = $E_\infty$ - $E_e$, and the position of the repulsive wall R$_{\textrm{in}}$ at the well depth. The relative change $\Delta$$E_\infty$ and $\Delta$$E_e$ of $E_\infty$ and $E_e$ with the increasing size of the basis set are also reported. Their progressions show a convergence similar to the one observed on the Li$^+$($^1$S$_0$) + Li($^2$S$_{1/2}$) (Table \ref{tab:tab1Li+Li}). 
The energy of Li$^+$($^1$S$_0$) + Li($^2$S$_{1/2}$) in the complete basis set (CBS) limit is the best variational representation of the dissociation limit, $E_\infty$, and should ideally be attained in the Full CI, (FCI), and CBS limit of the Li$_2^+$, X$^2\Sigma_g^+$ and A$^2\Sigma_u^+$ PECs. The difference between Li$^+$($^1$S$_0$) + Li($^2$S$_{1/2}$) obtained from the best available atomic calculation, listed in the Table I \cite{drake2006springer,yan1995eigenvalues}, and the $E_\infty$, obtained from aug-cc-pCV5Z calculation, listed in Table II, is smaller than the difference in the $E_\infty$ values obtained from the two cases X $\equiv$ Q and X $\equiv$ 5, suggesting a good convergence. The observed bound on the $E_\infty$ suggests that molecular calculations of the Li$_2^+$ in the FCI and CBS limits will not result in a change in the well depth, $D_e$, of X$^2\Sigma_g^+$ bigger than 10 cm$^{-1}$ (i.e. the difference between the $D_e$' values obtained in the X $\equiv$ Q and X $\equiv$ 5 cases) from the value obtained with the aug-cc-pCV5Z basis set. The experimental value of the $D_e$ (Table \ref{tab:tab3Li2+prev}), also supports the above theoretical bound.

The \textit{ab initio} X$^2\Sigma_g^+$ and A$^2\Sigma_u^+$ PECs, hereafter denoted as $\text{V}^{ab}_{g}$ and $\text{V}^{ab}_{u}$ respectively, relative to $E_\infty$, are shown in Fig.  \ref{fig:PECLi2+a}. The lower inset displays the difference between these PECs with the ones obtained from the method of \cite{aymar2005calculation} based on the representation of the Li$^+$ cores by an effective core potential (ECP) and a core polarization potential (CPP) (referred as the Th$_2$ method), thus treating the Li$_2^+$ molecule as a one-electron system (see also for instance \cite{magnier1999potential}). The overall agreement is satisfactory between the two approaches, with the largest difference at 12 a$_0$ of about 1\% in energy. Below 6 a$_0$, the difference is much larger, which can be understood as the ECP+CPP approach restrains the calculation from precisely representing the core-valence correlation at short internuclear distances.

\subsection{Determination of asymptotic extension of PECs} 
\label{sec:LR}

\begin{figure}[t]
	\includegraphics[width=0.490\textwidth, angle = 0]{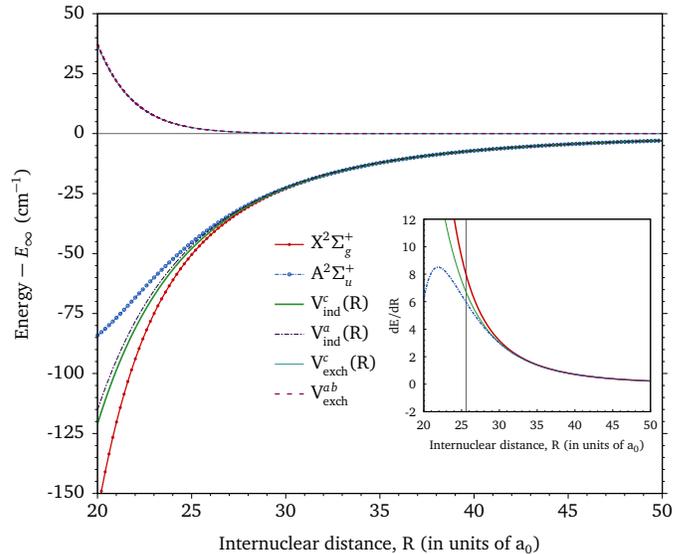}
	\caption{The asymptotically extended PECs  X$^2\Sigma_g^+$ (full red line) and A$^2\Sigma_u^+$ (full blue line) of $^7$Li$_2^{+}$ are shown. The asymptotic  induction function and \textit{ab initio }exchange term, V$^a_\text{ind}$(R) and V$^{ab}_{\text{exch}}$, the computed induction and exchange functions, V$^c_\text{ind}$(R) and V$^c_\text{exch}$(R), are plotted for the comparison. The first derivatives of X$^2\Sigma_g^+$ and A$^2\Sigma_u^+$ PECs, and V$^c_\text{ind}$(R) are drawn in the inset.}  
	\label{fig:PECLi2+b}
\end{figure}

\begin{figure*}[t]
\includegraphics[width=0.85\textwidth, angle = 0]{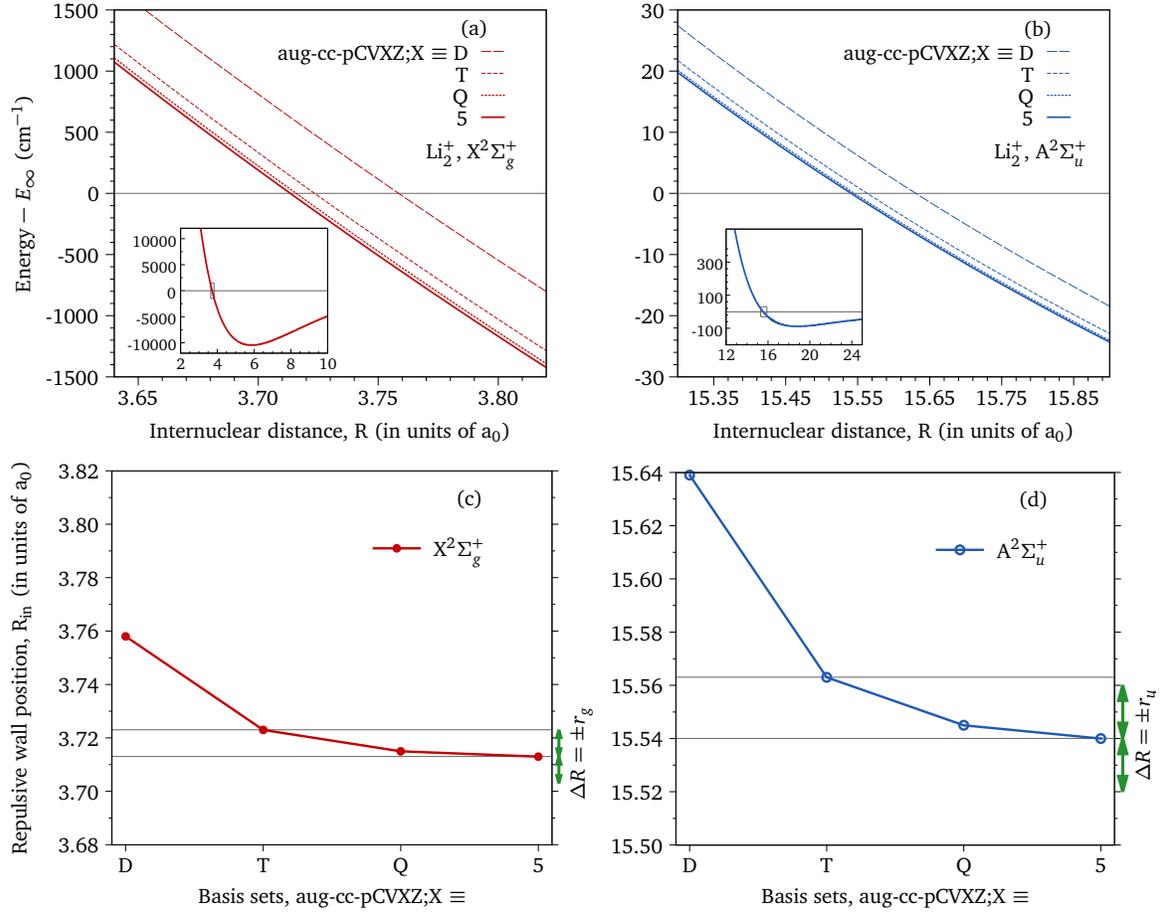}
\caption{Potential energy curves near the repulsive wall (see insets) for X$^2\Sigma_g^+$ (a) and A$^2\Sigma_u^+$ (b), computed using basis sets aug-cc-pCVXZ with X $\equiv$ D, T, Q, 5 and corresponding position R$_{\textrm{in}}$ of their inner turning point at the dissociation limit $E_\infty$ ((c) and (d)). The selected ranges $\Delta$R for the variation of the repulsive wall of the aug-cc-pCV5Z calculations mimicking possible inaccuracies for cross-section calculations are shown: $\Delta$R = $\pm$r$_{g/u}$ with r$_g$ = 0.01 a$_0$ and r$_u$ = 0.02 a$_0$ for X$^2\Sigma_g^+$ and A$^2\Sigma_u^+$ respectively.}
\label{fig:PECLi2+c}
\end{figure*}

The low-energy scattering wavefunctions need to be computed up to the large internuclear distances, with R $\gg\lambda$, where $\lambda$ is the de-Broglie wavelength of the colliding system (for $^7$Li$^+$-$^7$Li, 10 a$_0$ $<\lambda<$ 10$^6$ a$_0$ for the collision energies $10^{-5}$a.u. $>\text{E} > 10^{-15}$ a.u.).
The \textit{ab initio} PECs, in the large-R limit, generally become less accurate as the molecular orbitals which are built during the calculations are not best adapted to the situation of the separated atoms. Instead, we use the well-established asymptotic functional form $\text{V}^{a}_{p}\text{(R)}$ derived from the multipolar expansion of the interaction energy in inverse powers of R,

\begin{eqnarray}
\text{V}^{a}_{p}\text{(R)} = \text{V}^a_{\text{ind}}\text{(R)} \mp \text{V}^a_{\text{exch}}\text{(R)}; p \equiv \{g,u\},
\label{eq:eqLi2+1}
\end{eqnarray}
where $g$ (resp. $u$) corresponds to X$^2\Sigma_g^+$ (resp. A$^2\Sigma_u^+$).  The asymptotic induction term, V$^a_{\text{ind}}$(R) is expressed as \cite{cote2016ultracold} 
\begin{eqnarray}
\text{V}^a_{\text{ind}}\text{(R)} = -\bigg[~ \frac{C_4}{\text{R}^4} + \frac{C_6}{\text{R}^6} + \frac{C_8}{\text{R}^8} + ... ~\bigg],
\label{eq:eqLi2+2}
\end{eqnarray}
where $C_4$ = $\alpha_d/2$, $C_6$ = $\alpha_q/2$, $C_8$ = $\alpha_o/2$, with $\alpha_d$, $\alpha_q$, and $\alpha_o$ being the dipole, quadrupole, and octupole static polarizabilities of the $^7$Li ground state atom. We take the values from Tang \emph{et al.} \cite{tang2009nonrelativistic,mitroy2010theory},  $\alpha_d = 164.161$~a.u., $\alpha_q= 1423.415$~a.u., and $\alpha_o = 39653.720$~a.u. The van der Waals (dispersion) interaction, also varying as 1/R$^6$, which is generally small for ion-atom cases \cite{cote2016ultracold,zhang2009near}, will be included in an effective manner in the potential finally used in the scattering calculations.

The asymptotic exchange term reads \cite{bardsley1975calculations}
\begin{eqnarray}
\text{V}^a_{\text{exch}}\text{(R)} = \frac{1}{2}A \text{R}^{\alpha}e^{-\beta \text{R}} \Big[ 1 + \frac{B}{\text{R}} + \frac{C}{\text{R}^2} + ... \Big],
\label{eq:eqLi2+3}
\end{eqnarray}
where the parameters $\alpha = 2.1774$~a.u., $\beta = 0.6294$~a.u., and $B = 0.5191$~a.u. are simple functions of the $^7$Li ionization energy \cite{cote2016ultracold,bardsley1975calculations}. The $A$ and $C$ parameters are obtained from the fits of the \textit{ab initio} exchange energy, V$^{ab}_\text{exch}$, given by half of the difference of \textit{ab initio} A$^2\Sigma_u^+$ (V$^{ab}_{u}$) and X$^2\Sigma_g^+$ (V$^{ab}_{g}$) PECs with Eq. (\ref{eq:eqLi2+3}). The interval 23 a$_0$ $<$ R $<$ 28 a$_0$ is used in the fitting procedure, yielding $A = 0.133899$~a.u. and $C = 27.7397$~a.u. The selected interval gives us the fit with the smallest relative residuals. The same $A$ and $C$ provide excellent fit for the entire range above R $>$ 28 a$_0$. It suggests that for the exchange energy, the selected range represents the asymptotic limit, and it fixes the function form of exchange energy for $^7$Li$_2^+$, i.e. V$^c_\text{exch}$(R). The \textit{ab initio} exchange energy, V$^{ab}_\text{exch}$, intersects the function V$^c_\text{exch}$(R) at R = 25.6 a$_0$, which is selected as the point beyond which the asymptotic expansions are used. V$^a_\text{exch}$(R) (or V$^c_\text{exch}$(R)) decays exponentially with R, so in the large-R limit, only the contribution of V$^a_\text{ind}$(R) remains significant. Around 35 a$_0$, V$^a_\text{exch}$(R) becomes smaller than $0.1 \%$ of V$^a_\text{ind}$(R). Moreover, the contributions of the $C_6$/R$^6$ and $C_8$/R$^8$ terms become smaller than 1$\%$ of the induction energy beyond 29.5 a$_0$ and 12.5 a$_0$ respectively. In the internuclear range where only the $C_4$/R$^4$ term contributes significantly, $E_\infty$ is obtained using a fit on the \textit{ab initio} induction energy, given by average of A$^2\Sigma_u^+$ (V$^{ab}_{u}$) and X$^2\Sigma_g^+$ (V$^{ab}_{g}$) PECs, with the form given in the Eq. (2) using $C_6$ as a free parameter in the range 35-50 a$_0$. For a calculation with aug-cc-pCV5Z basis set, the change in the $E_\infty$ for different fit ranges, varying from 25 a$_0$-50 a$_0$ to 35 a$_0$-50 a$_0$, is only $\sim$ 0.02~cm$^{-1}$.

\setlength{\tabcolsep}{8.0pt}
\begin{table*}[]
\vspace{2pt}
\begin{tabular}{lcccccL{3.0cm}}
\hline
State and Method & $R_e$  & $D_e$ & $\omega_e$ & $\omega_e$$x_e$ & $B_e$  & Ref. \\
X$^2\Sigma_g^+$/A$^2\Sigma_u^+$ & (units of a$_0$) & (cm$^{-1}$) & (cm$^{-1}$) & (cm$^{-1}$) & (cm$^{-1}$) & No.\\
\hline
& & & X & & & \\
\hspace{12pt}$\text{Exp}$ & 5.88~ & 10464$\pm$6 & 262.2$\pm$1.5 & 1.7$\pm$0.5 & 0.496$\pm$0.002 & \cite{bernheim1984ionization,bernheim1983rydberg}\\

\hspace{12pt}$\text{Th$_1$}$ & 5.858 & 10458.58 & 261.96 & 1.51 & 0.500 & present study\\
\hspace{12pt}$\text{Th$_2$}$ & 5.838 & 10515.76 & 262.54 & 1.50 & 0.503 & present  study\\
\hspace{12pt}$\text{Th$_3$}$ & 5.863 & 10439 & 262.58 & 1.58 & -- & \cite{musial2015potential} \\
\hspace{12pt}$\text{Th$_4$}$ & 5.877 & 10457.7 & 261.6 & 1.47 & -- & \cite{zhang2009near} \\
\hspace{12pt}$\text{Th$_5$}$ & 5.844 & 10498 & 263.39 & -- & -- & \cite{jasik2007calculation} \\
\hspace{12pt}$\text{Th$_6$}$ & 5.848 & 10475 & 264 & 1.94 & 0.506 & \cite{bouzouita2006ab} \\
\hspace{12pt}$\text{Th$_7$}$ & 5.856 & 10441 & 263.76 & 1.646 & 0.5006 & \cite{schmidt1985ground} \\
\hspace{12pt}$\text{Th$_8$}$ & 5.826 & 10494 & 262.771 & 1.645 & 0.505 & \cite{rabli2017revised} \\
\hspace{12pt}$\text{Th$_9$}$ & 5.899 & 10466 & 263.08 & 1.477 & 0.4945 & \cite{magnier1999potential} \\
\hspace{12pt}$\text{Th$_{10}$}$ & 5.877 & 10457 & 266.2 & -- & 0.4753 & \cite{nasiri2017benchmark} \\
& & & A & & & \\
\hspace{12pt}$\text{Th$_1$}$ & 18.799 & 88.37 & 16.15 & 0.84 & 0.0486 & present study\\
\hspace{12pt}$\text{Th$_2$}$ & 18.797 & 88.71 & 16.17 & 0.84 & 0.0486 & present study\\
\hspace{12pt}$\text{Th$_3$}$ &  18.795 & 88  & 15.98 & 0.81 & -- & \cite{musial2015potential} \\
\hspace{12pt}$\text{Th$_4$}$ & 18.798 & 88.4 & 16.63 & 1.05 & -- & \cite{zhang2009near} \\
\hspace{12pt}$\text{Th$_5$}$ & 18.787 & 89 & 15.92 & -- & -- & \cite{jasik2007calculation} \\
\hspace{12pt}$\text{Th$_6$}$ & 18.729 & 88 & 15.81 & 0.74 & 0.049 & \cite{bouzouita2006ab} \\
\hspace{12pt}$\text{Th$_7$}$ & 18.802 & 90 & 20.1 & 0.13 & 0.049 & \cite{schmidt1985ground} \\
\hspace{12pt}$\text{Th$_8$}$ & 18.763 & 89 & 16.312 & 0.750 & 0.0487 & \cite{rabli2017revised} \\
\hspace{12pt}$\text{Th$_9$}$ & 18.899 & 90 & 16.01 & 0.79 & 0.049 & \cite{magnier1999potential} \\
\hline
\end{tabular}
\caption{Fundamental spectroscopic constants of the  X$^2\Sigma_g^+$ and A$^2\Sigma_u^+$ PECs for $^7$Li$_2^+$. The label Exp. refers to the best available experimental determination, while the numbered Th labels refer to various theoretical determinations.}
\label{tab:tab3Li2+prev}
\end{table*}

After setting $E_\infty$ as the origin of energies of the PECs, calculation of the extension of potentials for large R is performed under the following conditions: 
(i) the PECs X$^2\Sigma_g^+$ and A$^2\Sigma_u^+$ used in the scattering calculations and their derivatives are kept continuous at R = 25.6 a$_0$, (ii) the PECs approach V$^a_{p}$(R) as R $\rightarrow \infty$. First, a R-dependent  coefficient, $C_4$(R), is determined by expressing the \textit{ab initio} PECs in the range 20 a$_0 <$ R $<$ 50 a$_0$ as
\begin{eqnarray}
\text{V}_p^{ab} = \text{V}^c_{\text{ind}}\text{(R)} \mp \text{V}^c_{\text{exch}}\text{(R)}
\label{eq:eqLi2+c}
\end{eqnarray}
with
\begin{eqnarray}
 \text{V}^c_{\text{ind}}\text{(R)} = -\bigg[~\frac{C_4(\text{R})}{\text{R}^4}+\frac{C_6}{\text{R}^6} + \frac{C_8}{\text{R}^8} \bigg],
\label{eq:eqLi2+c-ind}
\end{eqnarray}
and functional form of V$^c_{\text{exch}}\text{(R)}$ which is determined previously. Then, from the computed $C_4$(R), functional forms of the $\partial C_4$/$\partial$R, and $C_4$(R), and consequently of V$^c_{\text{ind}}$(R) are obtained. In this way, the small van der Waals term is included in the function V$^c_{\text{ind}}$(R) in an effective way. The final scattering potentials X$^2\Sigma_g^+$ and A$^2\Sigma_u^+$ denoted as V$^c_{p}$(R) use \textit{ab initio} values for R $<$ 25.6 a$_0$ and V$^c_{\text{ind}}$(R) $\mp$ V$^c_{\text{exch}}$(R) for R $>$ 25.6 a$_0$.

The asymptotically extended PECs,  X$^2\Sigma_g^+$ and A$^2\Sigma_u^+$, V$^c_{p}$(R), the asymptotic induction function and \textit{ab initio} exchange energy, V$^a_{\text{ind}}$(R) and V$^{ab}_{\text{exch}}$, and the computed induction and exchange functions, V$^c_{\text{ind}}$(R) and V$^c_{\text{exch}}$(R), are shown in the Fig. \ref{fig:PECLi2+b}. The difference between V$^a_{\text{ind}}$(R), which uses a constant $C_4$, and V$^c_{\text{ind}}$(R), which uses a derived R-dependent function $C_4$(R), is quite evident in the 20-25 a$_0$ range (see Fig. \ref{fig:PECLi2+b}). This procedure fixes in a consistent way the asymptotic form of the PECs for reliable scattering calculations at extremely low energies.

\subsection{Criterion for bounds on the scattering parameters} 
\label{sec:SnB}

The previous section demonstrates that the asymptotic ion-atom interaction is well determined by the highly accurate calculations. Therefore the large variation in the low energy ion-atom cross section reported in the literature is illustrative of its strong sensitivity to the \textit{ab initio} part of the PECs given that the small-R region of the potentials is strongly influenced by the growing contribution of the core-electrons and thus is represented least accurately. 
To estimate the effect of this dependence on the scattering parameters, a set of PECs for X$^2\Sigma_g^+$  and A$^2\Sigma_u^+$ is generated by continuously varying the potentials according to 

\begin{eqnarray}
\text{R}^p = \text{R} + \text{r}_p( \text{R} - R_e)/(\text{R}_\text{{in}} - R_e) ~~~ \forall~ \text{R} < R_e,
\label{eq:repwall}
\end{eqnarray}

where \text{R}$^p$ denotes the co-ordinate of the generated PECs, and r$_p$ is the change in the repulsive wall position R$_\text{{in}}$. 
The allowed variation in the small-R region (i.e. R $<$ $R_e$) of the potentials is estimated by comparing the $D_e$ from the PECs obtained using different methods and basis sets (Table \ref{tab:tab3Li2+prev}), with our values computed with basis sets aug-cc-pCVXZ with X $\equiv$ D, T, Q, 5 (Table \ref{tab:tab2Li2+basis}). The difference in the well depths for the X$^2\Sigma_g^+$ obtained from aug-cc-pCVTZ and aug-cc-pCV5Z covers nearly similar variation, $\approx$ 40 cm$^{-1}$, observed from X$\text{Th$_{1-9}$}$ and X$_\text{Exp}$ (Table \ref{tab:tab3Li2+prev}). The difference between repulsive wall of PECs computed using aug-cc-pCV5Z and aug-cc-pCVTZ basis sets is thus taken as the permissible range of change in the wall positions of the PEC models with $\Delta$R = $\pm$r$_p$; $p\equiv \{g,u\}$ with r$_g$ = 0.01 a$_0$ for X$^2\Sigma_g^+$ and r$_u$ = 0.02 a$_0$ for A$^2\Sigma_u^+$. The determined energy bound for the allowed change in the small-R is much larger than the contributions arising from relativistic effects, diagonal Born-Oppenheimer correction (DBOC), and other corrections. A comparison is provided in the Section IV. 
The sets of PECs are created using the linear scaling of Eq. \ref{eq:repwall} for the required change of $\Delta$R = $\pm$r$_p$; $p\equiv \{g,u\}$ at the repulsive wall position R$_{\text{in}}$. The scattering calculations are performed for the two limiting modifications to both X$^2\Sigma_g^+$ and A$^2\Sigma_u^+$ curves with suffixes ":$\Delta$R = $\pm$r$_{g/u}$", and for \textit{ab initio} curves denoted as ":$\Delta$R = 0". 

An extensive comparison of the present results for the states, obtained with the aug-cc-pCV5Z basis set, X$^2\Sigma_g^+$:$\Delta$R = 0 and A$^2\Sigma_u^+$:$\Delta$R = 0, (referred to as Th$_1$), with those previously published in the literature is presented in Table \ref{tab:tab3Li2+prev}.
The vibrational levels of the X$^2\Sigma_g^+$:$\Delta$R = 0 and A$^2\Sigma_u^+$:$\Delta$R = 0 curves are evaluated using the LEVEL numerical code \cite{leroy2017level}. 
The X$^2\Sigma_g^+$ (resp. A$^2\Sigma_u^+$) PEC supports 82 (resp. 16) vibrational levels with vibrational harmonic constant $\omega_e = 261.96 $~cm$^{-1}$ and anharmonicity constant $\omega_ex_e$ = 1.51 cm$^{-1}$ (resp. $\omega_e = 16.15 $~cm$^{-1}$ and $\omega_ex_e$ = 0.84 cm$^{-1}$). The overall shape of the bottom of potential curve, described by $\omega_e$, $\omega_e x_e$ and $B_e$, is well reproduced by all calculations. They are in good agreement with the best available results from Optical-Optical Double Resonance, (OODR), spectroscopy \cite{bernheim1984ionization,bernheim1983rydberg}, falling within the reported error bars. We see that the present approach (Th$_1$) and the simpler method (Th$_2$), mentioned in Sec. II. A, are in remarkable agreement (about 0.5\% for the equilibrium distance $R_e$, the well depth $D_e$, and the rotational constant $B_e$, and even 0.2\% on the vibrational constant $\omega_e$).

Up to now the calculations are performed with the core-optimized basis set and the active core, \textit{i.e.} core excitations are included. To assess the contribution of the core electrons, we have performed an additional set of MRCI calculations with the cc-pVXZ;X $\equiv$ D, T, Q, 5, which are the basis sets self-consistently produced from the atomic calculations with the frozen-core electrons. The cc-pVXZ;X $\equiv$ D, T, Q, 5 basis sets are similar to the aug-cc-pCVXZ;X $\equiv$ D, T, Q, 5, the core-optimized basis with an augmented function, that are used in this work. PECs computed with cc-pVXZ;X $\equiv$ D, T, Q, 5 are mostly similar to their respective aug-cc-pCVXZ;X $\equiv$ D, T, Q, 5 PECs in the large-R region but are significantly inaccurate in the small-R region (especially R $<$ $R_e$). These calculations show that the repulsive wall positions of the PECs obtained from the frozen-core basis sets erroneously fall below the repulsive walls of the respective PECs with the active-core basis sets. In the case of frozen-core basis sets and frozen-core calculations, the unoptimized core continues to retain larger electron densities between the two nuclei than the cases when they are energy optimized along with the valence electrons. It, consequently, pushes the repulsive wall to the lower values in R much beyond the convergence limit shown in the Fig. 3 (c) and (d). We find that, for the small electronic systems, it is essential that PEC calculations are performed with the core-optimized basis sets in which all electrons of the molecular system are variationally optimized.

\section{$^7$Li$^+$-$^7$Li Collision cross sections}
\label{sec:cs}

\begin{figure}[t]
\includegraphics[width=0.490\textwidth, angle = 0]{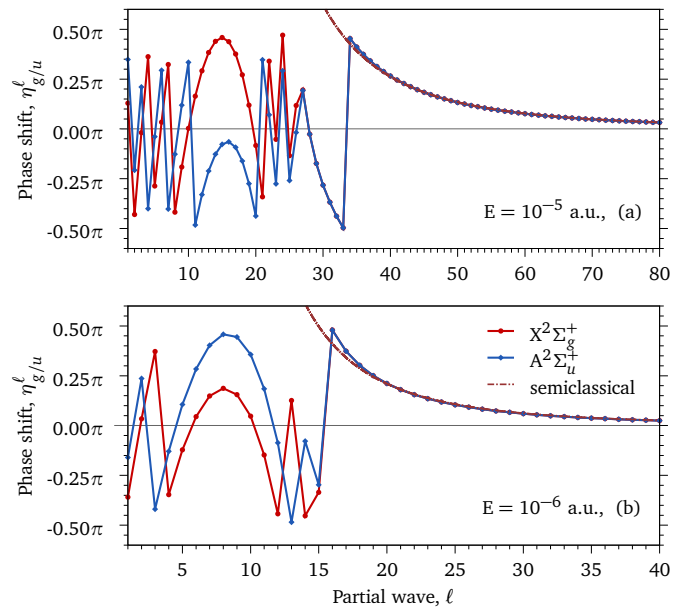}
\caption{Quantum (modulo $\pi$) and semiclassical phase shifts as functions of the partial waves, $\ell$, for a collision along the X$^2\Sigma_g^+$:$\Delta$R = 0 and A$^2\Sigma_u^+$:$\Delta$R = 0 curves for the collision energies (a) $10^{-5}$~a.u., and (b)  $10^{-6}$~a.u. The lines joining the points are a guide to the eye.}
\label{fig:Phshl}
\end{figure}

Applying standard scattering theory based on the partial wave expansion of the total wave function in R, the Schr\"{o}dinger equation for a single partial wave, $\ell$, at a collision energy E $=\hbar^2k^2/(2\mu)$; $k=2\pi/\lambda$ is  
\begin{eqnarray}
\Bigg[-\frac{\hbar^2}{2\mu}\frac{d^2}{d\text{R}^2} + \frac{\hbar^2}{2\mu}\frac{\ell(\ell+1)}{\text{R}^2} + \text{V}_{p}^c\text{(R)} \Bigg]y^{E,\ell}_{p}\text{(R)} ~~~~~\hspace{10pt}\nonumber \\= \text{E} y^{E,\ell}_{p}\text{(R)}, 
\label{eq:eqLi2+4}
\end{eqnarray}
where $\mu$ is the ($^7$Li$^+$-$^7$Li) Watson's charge-modified reduced mass \cite{watson1980isotope}. The asymptotic form of the wave function $y^{E,\ell}_{p}$(R) is  given by $y^{E,\ell}_{p}(\text R) \simeq k\text R [j_{\ell}(k\text R)cos(\eta^{\ell}_{p})- n_{\ell}(k\text R)sin(\eta^{\ell}_{p})]$, where $j_{\ell}(k\text R)$ and $n_{\ell}(k\text R)$ are the spherical Bessel functions, and $\eta^{\ell}_{p}$ is the quantum phase shift generated by the scattering potential V$^c_{p}$(R). Equation (\ref{eq:eqLi2+4}) is solved numerically, and $\eta^{\ell}_{p}$ is extracted at large distances, namely at R = $10\lambda$ as the asymptotic limit for low energies when $\lambda >$ 100~ a$_0$, and at R = 1000~ a$_0$ for higher energies when $\lambda <$ 100~ a$_0$.

In Fig.~\ref{fig:Phshl}, the quantum phase shifts $\eta^{\ell}_{p}$ (modulo $\pi$) are shown for E $= 10^{-5}$~a.u. (or $\sim 2 $ cm$^{-1}$), and E $= 10^{-6}$~a.u. (or $\sim 0.2$ cm$^{-1}$). For large $\ell$, when the outer classical turning point at a given collision energy, is such that V$^c_{p}$(R) can be approximated to the leading term $-\alpha_d$/2R$^{4}$ of V$^a_{\text{ind}}$(R), one can define the semiclassical phase shift as $\eta^{\ell}_{sc} \approx (\pi \mu^2 \alpha_d)/(4\hbar^4) \times \text{E}/{\ell}^3$ and thus the semiclassical cross section, $\sigma_{sc}$(E) = $\pi ( \mu \alpha_d^2/\hbar^2)^{1/3} (1 + \pi^2/16) \times \text{E}^{-1/3}$ \cite{cote2000ultracold}. The semiclassical phase shifts are in agreement with the quantum phase shifts for $\ell > L_{sc}$, with $L_{sc}=41$ for E $= 10^{-5}$~a.u. and $L_{sc}=19$ for E $= 10^{-6}$~a.u. Around E $ =10^{-8}$~a.u. (or $\sim 0.002$ cm$^{-1}$), as the contribution to the cross section from partial waves $\ell > 10$ becomes negligible, resonance features arise.

\begin{figure}[t]
\includegraphics[width=0.490\textwidth, angle = 0]{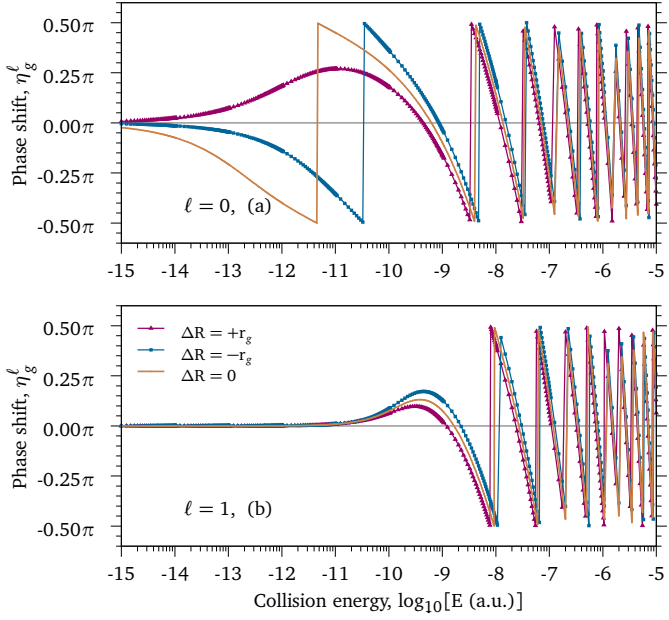}
\caption{Quantum phase shift (modulo $\pi$) of the X$^2\Sigma_g^+$:$\Delta$R = $\pm$r$_g$ and \textit{ab initio} X$^2\Sigma_g^+$:$\Delta$R = 0 curves as a function of the collision energy for the partial waves  (a) $\ell=0$, and (b) $\ell=1$. At low energies, the change in the phase shifts for different PEC models are significant only for $\ell$ = 0.}
\label{fig:Phshe}
\end{figure}

In Fig. \ref{fig:Phshe}, the quantum phase shifts $\eta^{\ell}_g$ for ${\ell}= 0, 1$  are plotted as a function of the collision energy for the \textit{ab initio} PEC, X$^2\Sigma_g^+$:$\Delta$R = 0, and the generated PECs with shifted repulsive walls, X$^2\Sigma_g^+$:$\Delta$R = $\pm$r$_g$. At low energies, the effect is weak for $\ell > 0$ as the centrifugal barrier becomes dominant in the collision. Note that the $s$-wave ($\ell=0$) phase shift changes sign when the repulsive wall is slightly shifted, indicating the presence of a pole where the scattering length diverges. As a result, the accuracy of the PEC becomes a major factor in determining the collision cross section. This is the primary motivation for the extreme care taken in determining the scattering potential in Section II.

\begin{figure*}[t]
\includegraphics[width=0.995\textwidth, angle = 0]{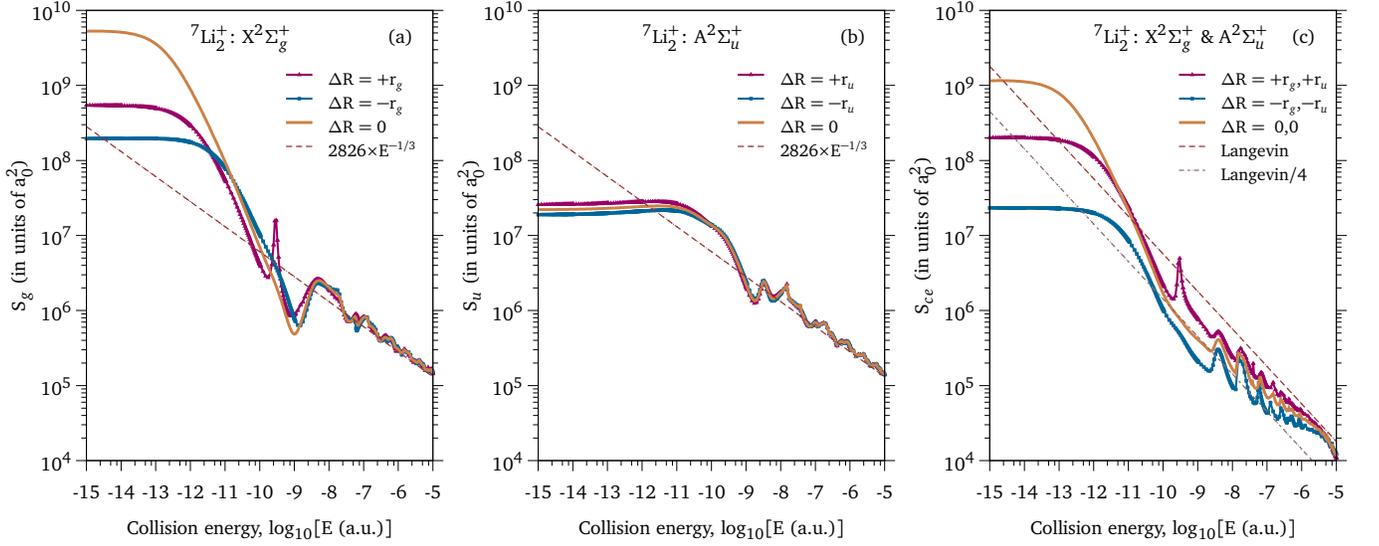}
\caption{(a) S$_{g}$(E) for the generated X$^2\Sigma_g^+$:$\Delta$R = $\pm$r$_g$, and X$^2\Sigma_g^+$:$\Delta$R = 0 curves are plotted. (b) S$_{u}$(E) for the A$^2\Sigma_u^+$:$\Delta$R = $\pm$r$_u$  and A$^2\Sigma_u^+$:$\Delta$R = 0 are plotted. In (a),(b), the semiclassical cross section, 2826$\times$E$^{-1/3}$, is shown. (c) S$_{ce}$(E) for the two bounding modifications of PECs, X$^2\Sigma_g^+$:$\Delta$R = $+$r$_g$, A$^2\Sigma_u^+$:$\Delta$R = $+$r$_u$  and X$^2\Sigma_g^+$:$\Delta$R = $-$r$_g$, A$^2\Sigma_u^+$:$\Delta$R = $-$r$_u$, are shown along with the S$_{ce}$(E) for X$^2\Sigma_g^+$:$\Delta$R = 0, A$^2\Sigma_u^+$:$\Delta$R = 0 curves. Langevin and Langevin/4 are also plotted for comparison.}
\label{fig:elsgsuexch}
\end{figure*}

Due to the identical nuclei, the scattering between $^7$Li$^+$-$^7$Li ion-atom system enables the event in which the ion and the atom exchange their charge identities. A scattering event when the initial identities are preserved is a direct elastic collision, whereas the event when the identities of the ion-atom pair are interchanged is termed as resonant charge exchange collision (RCE) \cite{massey1972atom}. The scattering amplitudes for direct elastic and RCE collisions are given by ($f_g$ + $f_u$)/2 and $f_{ce}$ = ($f_g$ - $f_u$)/2 where $f_g$ and $f_u$ are scattering amplitudes for X$^2\Sigma_g^+$ and A$^2\Sigma_u^+$. We define S$_g$(E) and S$_u$(E) in Eq. \ref{eq:eqLi2+5} and S$_{ce}$(E) in Eq. \ref{eq:eqLi2+6}, where d$\Omega$ is the differential solid angle, as

\begin{eqnarray}
\text{S}_{p}\text{(E)} = \int |f_{p}|^2 d\Omega = \frac{4 \pi}{k^2} \sum_{\ell=0}^{\infty} (2\ell +1) sin^2(\eta^\ell_{p}),
\label{eq:eqLi2+5} 
\end{eqnarray}

\begin{eqnarray}
\text{S}_{ce}\text{(E)} = \int |f_{ce}|^2 d\Omega = \frac{\pi}{k^2} \sum_{\ell=0}^{\infty} (2\ell+1) sin^2(\eta^\ell_{g}-\eta^\ell_{u}).
\label{eq:eqLi2+6}
\end{eqnarray}

The average (S$_{g}$(E) + S$_{u}$(E))/2 has been identified as the total cross section, and S$_{ce}$(E) as the RCE cross section when certain approximations are made \cite{cote2016ultracold} at high collision energies. The functions S$_{g}$(E) and S$_{u}$(E) for the $^7$Li$^+$-$^7$Li system as functions of the collision energy are shown along with the semiclassical scattering cross section, $\sigma_{sc}$(E), in Fig. \ref{fig:elsgsuexch} (a) and (b). For $^7$Li$^+$-$^7$Li, $\sigma_{sc}$(E) = 2826$\times$E$^{-1/3}$ a.u. The Langevin cross section, ( $ \sim \pi (2\alpha_d)^{1/2}\times \text{E}^{-1/2}$), for $^7$Li$^+$-$^7$Li, 56.92$\times$E$^{-1/2}$ a.u., and Langevin/4 are shown along with S$_{ce}$(E) in Fig. \ref{fig:elsgsuexch} (c). 
In all cases, cross sections include the sum of first 100 partial waves. It can be seen that S$_{ce}$(E), in this case, predominantly falls in the range defined by Langevin and Langevin/4. For low energies, S$_{ce}$(E) varies significantly from the expected semiclassical picture.

\begin{figure}[b]
\includegraphics[width=0.490\textwidth, angle = 0]{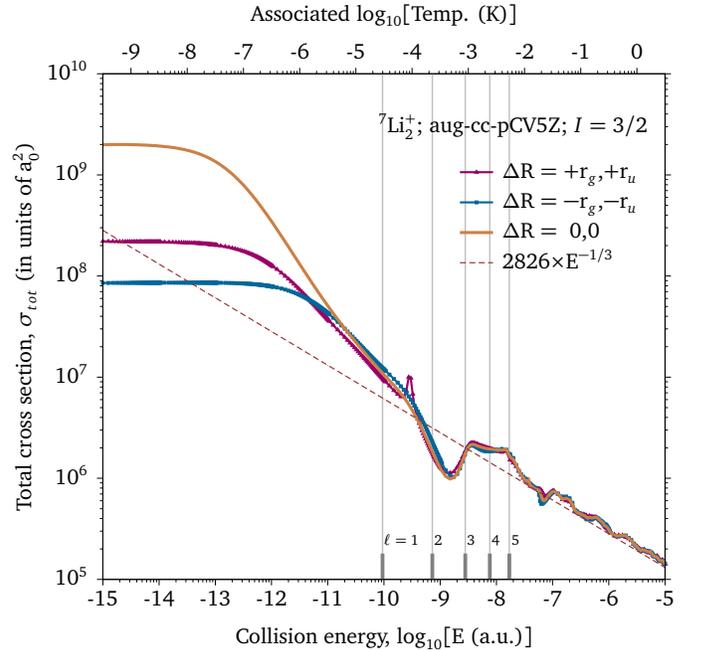}
\caption{The total collision cross section, $\sigma_{tot}$(E), of the $^7$Li$^+$-$^7$Li system in its first asymptotic state, which involves the electronic states X$^2\Sigma_g^+$ and A$^2\Sigma_u^+$ are shown for the modified PECs X/A:$\Delta$R = $\pm$(r$_{g}$,r$_{u}$), and X/A:$\Delta$R = 0,0. The semiclassical cross section, 2826$\times$E$^{-1/3}$, and the centrifugal barrier energies for $\ell$ = 1-5 are also shown.}
\label{fig:ttlcr}
\end{figure}

For homonuclear systems, in principle, individual scattering channels cannot be measured independently and therefore we compute the total cross section $\sigma_{tot}$(E), given in Eq. \ref{eq:eqLi2+7}. The expression for $\sigma_{tot}$(E) differs from the one usually employed in the literature; the derivation will be discussed elsewhere \cite{Nishant2019homonuclear}.

\begin{eqnarray}
\sigma_{tot}\text{(E)} = \hspace{200pt} \nonumber \\
\frac{4 \pi}{k^2}\Bigg[ x\Big[\sum_{even}^{}(2\ell+1)sin^2(\eta^\ell_{g})+\sum_{odd}^{} (2\ell +1) sin^2(\eta^\ell_{u})\Big] + \hspace{10pt}\nonumber \\
(1-x)\Big[\sum_{odd}^{}(2\ell +1)sin^2(\eta^\ell_{g})+\sum_{even}^{} (2\ell +1) sin^2(\eta^\ell_{u})\Big] \Bigg], \hspace{11pt}
\label{eq:eqLi2+7}
\end{eqnarray}

where $x$ is a function of the nuclear spin $I$. For a half-integer nuclear spin, $x$ = $I$/(2$I$+1). For $^7$Li, with $I$ = 3/2, $x$ is 3/8 \cite{cote2016ultracold}. The cross section evaluated using Eq. \ref{eq:eqLi2+7} differs significantly in the $s$-wave limit with the value calculated as the average of S$_{g}$(E) and S$_{u}$(E). For $^7$Li$^+$-$^7$Li, in the $s$-wave limit, cross section obtained using $\sigma_{tot}$(E) is 25$\%$ smaller than the average of S$_{g}$(E) and S$_{u}$(E).
The total cross section, $\sigma_{tot}$(E), for the $^7$Li$^+$-$^7$Li is plotted along with the semiclassical scattering cross section, $\sigma_{sc}$(E), in Fig. \ref{fig:ttlcr}. The centrifugal barrier energies introduced by the first few partial waves are also shown.

The scattering length $a_u$, when compared with the characteristic interaction length scale R$^\star$, i.e. position of the $\ell$ = 1 barrier $({\alpha_d \times \mu/\hbar^2)}^{1/2}$, which for $^7$Li$^+$-$^7$Li is 1024 a$_0$,  is within a factor of two, while $a_g$ is very large, see Table IV. Also, S$_g$(E) and $a_g$ are more sensitive to the small-R region of the PEC and consequently to the details of the short-range interaction than S$_u$(E) and $a_u$. This sensitivity for the X$^2\Sigma_g^+$ is amplified for $^7$Li$^+$-$^7$Li system, which is also noted by Schmid \emph{et al.} \cite{schmid2018rydberg}, due to proximity of a scattering pole, i.e. the PEC either is about to acquire or just acquired a weakly-bound state.

\setlength{\tabcolsep}{6.25pt}
\begin{table}[t]
	\vspace{2pt}
	\begin{tabular}{ccccc} 
		\hline
		X$^2\Sigma_g^+$, A$^2\Sigma_u^+$ & $\Delta$R = $\pm$r$_{g,u}$ & $\Delta$R = 0 & \cite{zhang2009near} & \cite{schmid2018rydberg}\\
		\hline
		 $a_{g}$ &  -6582$/$3948 & 20465 & 14337 & 7162  \\
		 $a_{u}$ &   1432$/$1227 & 1325 & 1262 & -- \\
		\hline
	\end{tabular}
	\caption{\label{tab:tab4Li2+scat} $^7$Li$^+$-$^7$Li scattering lengths for the modeled X$^2\Sigma_g^+$:$\Delta$R = $\pm$r$_g$, X$^2\Sigma_g^+$:$\Delta$R = 0 and A$^2\Sigma_u^+$:$\Delta$R = $\pm$r$_u$, A$^2\Sigma_u^+$:$\Delta$R = 0 curves are listed. For direct comparison with Zhang \emph{et al.} \cite {zhang2009near} and Schmid \emph{et al.} \cite{schmid2018rydberg}, the values obtained from X/A:$\Delta$R = 0 are appropriate.}
\end{table}

\section{Discussion and Conclusion}

The values of $D_e$ calculated by Zhang \textit{et al.} \cite{zhang2009near} and Schmid \textit{et al.} \cite{schmid2018rydberg} along with the value calculated in this work (Table III), fall within the experimental accuracy of 10464$\pm$6 cm$^{-1}$ \cite{bernheim1984ionization}. However, the convergence of $E_e$ and $E_\infty$, and the variational nature of the calculation provide an additional certainty in our case. We have calculated the relativistic corrections using second order Douglas-Kroll-Hess Hamiltonian \cite{MOLPRO-WIREs}. Relativistic corrections on the PECs can be expressed in two parts -- a constant shift by $\approx$ -306.6 cm$^{-1}$, and a R dependent change in the total electronic energy. The constant shift due to relativistic corrections does not affect the scattering calculations. The R dependent change in the total electronic energy is less than 1.0 cm$^{-1}$ for R $>$ $R_e$ and less than 5.0 cm$^{-1}$ for R $<$ $R_e$, which is not significant when compared with the effect of core-electrons in the calculation, which is $\pm$140 cm$^{-1}$ at the repulsive wall position, R$_{\textrm{in}}$, for the allowed change of $\pm$r$_g$ in the X$^2\Sigma_g^+$ curve. We have found that the variation in the DBOC is less than 0.5 cm$^{-1}$ in the entire internuclear range \cite{lutz2016deviations}. In addition, as we have discussed before, counterpoise corrections for BSSE is not relevant in our case.

In the present work, an analysis is performed to obtain consistent asymptotic extension of the scattering potentials. We find that $^7$Li$^+$-$^7$Li system in the X$^2\Sigma_g^+$ state is close to a scattering pole, and therefore extreme care is required in the computation of low energy scattering parameters. Scattering lengths for X$^2\Sigma_g^+$:$\Delta$R = 0 and A$^2\Sigma_u^+$:$\Delta$R = 0 are 20465 a$_0$ and 1325 a$_0$ respectively, ( see Table \ref{tab:tab4Li2+scat}). Scattering length, $a_g$, reported by Zhang \emph{et al.} \cite {zhang2009near} and Schmid \emph{et al.} \cite{schmid2018rydberg} are 14337 and 7162 a$_0$ respectively. Schmid \emph{et al.} also provides a bound on $a_g$ as (107825 a$_0$; 3664 a$_0$) that corresponds to the potentials scaled by (0.999; 1.001) to the computed PEC.
The possible errors in the cross section, in our case,  are estimated by controlled variations in the small-R region of the PECs, assessing the change they bring to the phase shifts and cross sections in the low energy limit. 
The scattering pole for X$^2\Sigma_g^+$ occurs within the determined range of variations as shown in the Fig. \ref{fig:PECLi2+c}, particularly in between the PEC models $\Delta$R = $+$r$_g$ and $\Delta$R = 0 which is also evident in the phase shift plot, Fig. \ref{fig:Phshe}, which prevents us from estimating the upper limit of the total cross-section. However, the lower limit of the total cross-section is given by the $\Delta$R = $-$r$_g$,$-$r$_u$ curve.  The setting of this range will prevent the values reported here to be affected by even more sophisticated calculations in the future. The calculated value of the total cross sections is shown by the $\Delta$R = 0 curve in Fig. \ref{fig:ttlcr}. The cross sections are determined for a wide range of collision energy, from 10$^{-5}$ to 10$^{-15}$ a.u., which covers a large range of temperatures from few K to few nK. 
S$_{g}$(E), S$_{u}$(E), S$_{ce}$(E), and $\sigma_{tot}$(E) in the temperature regimes below a few mK, have contributions only from few partial waves (about 5). In this regime, the cross sections significantly deviate from the semiclassical values and result in the distinctive features that can be explored in the future experiments. The total cross section for $^7$Li$^+$-$^7$Li system in the low energy limit is 1.9$\times 10^9$ a$_0^2$. 
When the collision energy is larger than a few mK, many partial waves participate in the scattering and their contributions sum up to give the semiclassical value.

\section{Acknowledgements}
The authors acknowledge support from IFCPAR/CEFIPRA grant No. 5404-1.

\bibliographystyle{apsrev}
\bibliography{references1st-mdf-arx.bib}

\begin{thebibliography}{51}
\expandafter\ifx\csname natexlab\endcsname\relax\def\natexlab#1{#1}\fi
\expandafter\ifx\csname bibnamefont\endcsname\relax
  \def\bibnamefont#1{#1}\fi
\expandafter\ifx\csname bibfnamefont\endcsname\relax
  \def\bibfnamefont#1{#1}\fi
\expandafter\ifx\csname citenamefont\endcsname\relax
  \def\citenamefont#1{#1}\fi
\expandafter\ifx\csname url\endcsname\relax
  \def\url#1{\texttt{#1}}\fi
\expandafter\ifx\csname urlprefix\endcsname\relax\def\urlprefix{URL }\fi
\providecommand{\bibinfo}[2]{#2}
\providecommand{\eprint}[2][]{\url{#2}}

\bibitem[{\citenamefont{Grier et~al.}(2009)\citenamefont{Grier, Cetina,
  Oru{\v{c}}evi{\'c}, and Vuleti{\'c}}}]{grier2009observation}
\bibinfo{author}{\bibfnamefont{A.~T.} \bibnamefont{Grier}},
  \bibinfo{author}{\bibfnamefont{M.}~\bibnamefont{Cetina}},
  \bibinfo{author}{\bibfnamefont{F.}~\bibnamefont{Oru{\v{c}}evi{\'c}}},
  \bibnamefont{and}
  \bibinfo{author}{\bibfnamefont{V.}~\bibnamefont{Vuleti{\'c}}},
  \bibinfo{journal}{Phys. Rev. Lett.} \textbf{\bibinfo{volume}{102}},
  \bibinfo{pages}{223201} (\bibinfo{year}{2009}).

\bibitem[{\citenamefont{Zipkes et~al.}(2010)\citenamefont{Zipkes, Palzer, Sias,
  and K{\"o}hl}}]{zipkes2010trapped}
\bibinfo{author}{\bibfnamefont{C.}~\bibnamefont{Zipkes}},
  \bibinfo{author}{\bibfnamefont{S.}~\bibnamefont{Palzer}},
  \bibinfo{author}{\bibfnamefont{C.}~\bibnamefont{Sias}}, \bibnamefont{and}
  \bibinfo{author}{\bibfnamefont{M.}~\bibnamefont{K{\"o}hl}},
  \bibinfo{journal}{Nature} \textbf{\bibinfo{volume}{464}},
  \bibinfo{pages}{388} (\bibinfo{year}{2010}).

\bibitem[{\citenamefont{Schmid et~al.}(2010)\citenamefont{Schmid, H{\"a}rter,
  and Denschlag}}]{schmid2010dynamics}
\bibinfo{author}{\bibfnamefont{S.}~\bibnamefont{Schmid}},
  \bibinfo{author}{\bibfnamefont{A.}~\bibnamefont{H{\"a}rter}},
  \bibnamefont{and} \bibinfo{author}{\bibfnamefont{J.~H.}
  \bibnamefont{Denschlag}}, \bibinfo{journal}{Phys. Rev. Lett.}
  \textbf{\bibinfo{volume}{105}}, \bibinfo{pages}{133202}
  (\bibinfo{year}{2010}).

\bibitem[{\citenamefont{Meir et~al.}(2016)\citenamefont{Meir, Sikorsky,
  Ben-Shlomi, Akerman, Dallal, and Ozeri}}]{meir2016dynamics}
\bibinfo{author}{\bibfnamefont{Z.}~\bibnamefont{Meir}},
  \bibinfo{author}{\bibfnamefont{T.}~\bibnamefont{Sikorsky}},
  \bibinfo{author}{\bibfnamefont{R.}~\bibnamefont{Ben-Shlomi}},
  \bibinfo{author}{\bibfnamefont{N.}~\bibnamefont{Akerman}},
  \bibinfo{author}{\bibfnamefont{Y.}~\bibnamefont{Dallal}}, \bibnamefont{and}
  \bibinfo{author}{\bibfnamefont{R.}~\bibnamefont{Ozeri}},
  \bibinfo{journal}{Phys. Rev. Lett.} \textbf{\bibinfo{volume}{117}},
  \bibinfo{pages}{243401} (\bibinfo{year}{2016}).

\bibitem[{\citenamefont{Ratschbacher et~al.}(2012)\citenamefont{Ratschbacher,
  Zipkes, Sias, and K{\"o}hl}}]{ratschbacher2012controlling}
\bibinfo{author}{\bibfnamefont{L.}~\bibnamefont{Ratschbacher}},
  \bibinfo{author}{\bibfnamefont{C.}~\bibnamefont{Zipkes}},
  \bibinfo{author}{\bibfnamefont{C.}~\bibnamefont{Sias}}, \bibnamefont{and}
  \bibinfo{author}{\bibfnamefont{M.}~\bibnamefont{K{\"o}hl}},
  \bibinfo{journal}{Nat. Phys.} \textbf{\bibinfo{volume}{8}},
  \bibinfo{pages}{649} (\bibinfo{year}{2012}).

\bibitem[{\citenamefont{Haze et~al.}(2013)\citenamefont{Haze, Hata, Fujinaga,
  and Mukaiyama}}]{haze2013observation}
\bibinfo{author}{\bibfnamefont{S.}~\bibnamefont{Haze}},
  \bibinfo{author}{\bibfnamefont{S.}~\bibnamefont{Hata}},
  \bibinfo{author}{\bibfnamefont{M.}~\bibnamefont{Fujinaga}}, \bibnamefont{and}
  \bibinfo{author}{\bibfnamefont{T.}~\bibnamefont{Mukaiyama}},
  \bibinfo{journal}{Phys. Rev. A} \textbf{\bibinfo{volume}{87}},
  \bibinfo{pages}{052715} (\bibinfo{year}{2013}).

\bibitem[{\citenamefont{Haze et~al.}(2015)\citenamefont{Haze, Saito, Fujinaga,
  and Mukaiyama}}]{haze2015charge}
\bibinfo{author}{\bibfnamefont{S.}~\bibnamefont{Haze}},
  \bibinfo{author}{\bibfnamefont{R.}~\bibnamefont{Saito}},
  \bibinfo{author}{\bibfnamefont{M.}~\bibnamefont{Fujinaga}}, \bibnamefont{and}
  \bibinfo{author}{\bibfnamefont{T.}~\bibnamefont{Mukaiyama}},
  \bibinfo{journal}{Phys. Rev. A} \textbf{\bibinfo{volume}{91}},
  \bibinfo{pages}{032709} (\bibinfo{year}{2015}).

\bibitem[{\citenamefont{Ravi et~al.}(2012)\citenamefont{Ravi, Lee, Sharma,
  Werth, and Rangwala}}]{ravi2012cooling}
\bibinfo{author}{\bibfnamefont{K.}~\bibnamefont{Ravi}},
  \bibinfo{author}{\bibfnamefont{S.}~\bibnamefont{Lee}},
  \bibinfo{author}{\bibfnamefont{A.}~\bibnamefont{Sharma}},
  \bibinfo{author}{\bibfnamefont{G.}~\bibnamefont{Werth}}, \bibnamefont{and}
  \bibinfo{author}{\bibfnamefont{S.}~\bibnamefont{Rangwala}},
  \bibinfo{journal}{Nat. Commun.} \textbf{\bibinfo{volume}{3}},
  \bibinfo{pages}{1126} (\bibinfo{year}{2012}).

\bibitem[{\citenamefont{Sivarajah et~al.}(2012)\citenamefont{Sivarajah,
  Goodman, Wells, Narducci, and Smith}}]{sivarajah2012evidence}
\bibinfo{author}{\bibfnamefont{I.}~\bibnamefont{Sivarajah}},
  \bibinfo{author}{\bibfnamefont{D.}~\bibnamefont{Goodman}},
  \bibinfo{author}{\bibfnamefont{J.}~\bibnamefont{Wells}},
  \bibinfo{author}{\bibfnamefont{F.}~\bibnamefont{Narducci}}, \bibnamefont{and}
  \bibinfo{author}{\bibfnamefont{W.}~\bibnamefont{Smith}},
  \bibinfo{journal}{Phys. Rev. A} \textbf{\bibinfo{volume}{86}},
  \bibinfo{pages}{063419} (\bibinfo{year}{2012}).

\bibitem[{\citenamefont{Dutta et~al.}(2017)\citenamefont{Dutta, Sawant, and
  Rangwala}}]{dutta2017collisional}
\bibinfo{author}{\bibfnamefont{S.}~\bibnamefont{Dutta}},
  \bibinfo{author}{\bibfnamefont{R.}~\bibnamefont{Sawant}}, \bibnamefont{and}
  \bibinfo{author}{\bibfnamefont{S.}~\bibnamefont{Rangwala}},
  \bibinfo{journal}{Phys. Rev. Lett.} \textbf{\bibinfo{volume}{118}},
  \bibinfo{pages}{113401} (\bibinfo{year}{2017}).

\bibitem[{\citenamefont{Hall et~al.}(2011)\citenamefont{Hall, Aymar,
  Bouloufa-Maafa, Dulieu, and Willitsch}}]{hall2011light}
\bibinfo{author}{\bibfnamefont{F.~H.} \bibnamefont{Hall}},
  \bibinfo{author}{\bibfnamefont{M.}~\bibnamefont{Aymar}},
  \bibinfo{author}{\bibfnamefont{N.}~\bibnamefont{Bouloufa-Maafa}},
  \bibinfo{author}{\bibfnamefont{O.}~\bibnamefont{Dulieu}}, \bibnamefont{and}
  \bibinfo{author}{\bibfnamefont{S.}~\bibnamefont{Willitsch}},
  \bibinfo{journal}{Phys. Rev. Lett.} \textbf{\bibinfo{volume}{107}},
  \bibinfo{pages}{243202} (\bibinfo{year}{2011}).

\bibitem[{\citenamefont{Schowalter et~al.}(2016)\citenamefont{Schowalter,
  Dunning, Chen, Puri, Schneider, and Hudson}}]{schowalter2016blue}
\bibinfo{author}{\bibfnamefont{S.~J.} \bibnamefont{Schowalter}},
  \bibinfo{author}{\bibfnamefont{A.~J.} \bibnamefont{Dunning}},
  \bibinfo{author}{\bibfnamefont{K.}~\bibnamefont{Chen}},
  \bibinfo{author}{\bibfnamefont{P.}~\bibnamefont{Puri}},
  \bibinfo{author}{\bibfnamefont{C.}~\bibnamefont{Schneider}},
  \bibnamefont{and} \bibinfo{author}{\bibfnamefont{E.~R.}
  \bibnamefont{Hudson}}, \bibinfo{journal}{Nat. Commun.}
  \textbf{\bibinfo{volume}{7}}, \bibinfo{pages}{12448} (\bibinfo{year}{2016}).

\bibitem[{\citenamefont{H{\"a}rter et~al.}(2012)\citenamefont{H{\"a}rter,
  Kr{\"u}kow, Brunner, Schnitzler, Schmid, and Denschlag}}]{harter2012single}
\bibinfo{author}{\bibfnamefont{A.}~\bibnamefont{H{\"a}rter}},
  \bibinfo{author}{\bibfnamefont{A.}~\bibnamefont{Kr{\"u}kow}},
  \bibinfo{author}{\bibfnamefont{A.}~\bibnamefont{Brunner}},
  \bibinfo{author}{\bibfnamefont{W.}~\bibnamefont{Schnitzler}},
  \bibinfo{author}{\bibfnamefont{S.}~\bibnamefont{Schmid}}, \bibnamefont{and}
  \bibinfo{author}{\bibfnamefont{J.~H.} \bibnamefont{Denschlag}},
  \bibinfo{journal}{Phys. Rev. Lett.} \textbf{\bibinfo{volume}{109}},
  \bibinfo{pages}{123201} (\bibinfo{year}{2012}).

\bibitem[{\citenamefont{Morita et~al.}(2018)\citenamefont{Morita, Sikorsky,
  Meir, Buchachenko, Ben-shlomi, Akerman, Narevicius, V~Tscherbul, and
  Ozeri}}]{morita2018spin}
\bibinfo{author}{\bibfnamefont{M.}~\bibnamefont{Morita}},
  \bibinfo{author}{\bibfnamefont{T.}~\bibnamefont{Sikorsky}},
  \bibinfo{author}{\bibfnamefont{Z.}~\bibnamefont{Meir}},
  \bibinfo{author}{\bibfnamefont{A.}~\bibnamefont{Buchachenko}},
  \bibinfo{author}{\bibfnamefont{R.}~\bibnamefont{Ben-shlomi}},
  \bibinfo{author}{\bibfnamefont{N.}~\bibnamefont{Akerman}},
  \bibinfo{author}{\bibfnamefont{E.}~\bibnamefont{Narevicius}},
  \bibinfo{author}{\bibfnamefont{T.}~\bibnamefont{V~Tscherbul}},
  \bibnamefont{and} \bibinfo{author}{\bibfnamefont{R.}~\bibnamefont{Ozeri}},
  \bibinfo{journal}{Bulletin of the American Physical Society}
  \textbf{\bibinfo{volume}{63}}, \bibinfo{pages}{H08.005}
  (\bibinfo{year}{2018}).

\bibitem[{\citenamefont{Sikorsky et~al.}(2018)\citenamefont{Sikorsky, Meir,
  Ben-Shlomi, Akerman, and Ozeri}}]{sikorsky2018spin}
\bibinfo{author}{\bibfnamefont{T.}~\bibnamefont{Sikorsky}},
  \bibinfo{author}{\bibfnamefont{Z.}~\bibnamefont{Meir}},
  \bibinfo{author}{\bibfnamefont{R.}~\bibnamefont{Ben-Shlomi}},
  \bibinfo{author}{\bibfnamefont{N.}~\bibnamefont{Akerman}}, \bibnamefont{and}
  \bibinfo{author}{\bibfnamefont{R.}~\bibnamefont{Ozeri}},
  \bibinfo{journal}{Nat. Commun.} \textbf{\bibinfo{volume}{9}},
  \bibinfo{pages}{920} (\bibinfo{year}{2018}).

\bibitem[{\citenamefont{Wang et~al.}(2019)\citenamefont{Wang, Dei{\ss},
  Raithel, and Denschlag}}]{wang2019controlling}
\bibinfo{author}{\bibfnamefont{L.}~\bibnamefont{Wang}},
  \bibinfo{author}{\bibfnamefont{M.}~\bibnamefont{Dei{\ss}}},
  \bibinfo{author}{\bibfnamefont{G.}~\bibnamefont{Raithel}}, \bibnamefont{and}
  \bibinfo{author}{\bibfnamefont{J.~H.} \bibnamefont{Denschlag}},
  \bibinfo{journal}{arXiv preprint arXiv:1901.08781}  (\bibinfo{year}{2019}),
  \eprint{1901.08781}.

\bibitem[{\citenamefont{Massey}(1972)}]{massey1972atom}
\bibinfo{author}{\bibfnamefont{H.}~\bibnamefont{Massey}},
  \bibinfo{journal}{Contemp. Phys.} \textbf{\bibinfo{volume}{13}},
  \bibinfo{pages}{135} (\bibinfo{year}{1972}).

\bibitem[{\citenamefont{Cetina et~al.}(2012)\citenamefont{Cetina, Grier, and
  Vuleti{\'c}}}]{cetina2012micromotion}
\bibinfo{author}{\bibfnamefont{M.}~\bibnamefont{Cetina}},
  \bibinfo{author}{\bibfnamefont{A.~T.} \bibnamefont{Grier}}, \bibnamefont{and}
  \bibinfo{author}{\bibfnamefont{V.}~\bibnamefont{Vuleti{\'c}}},
  \bibinfo{journal}{Phys. Rev. Lett.} \textbf{\bibinfo{volume}{109}},
  \bibinfo{pages}{253201} (\bibinfo{year}{2012}).

\bibitem[{\citenamefont{Feldker et~al.}(2020)\citenamefont{Feldker, F{\"u}rst,
  Hirzler, Ewald, Mazzanti, Wiater, Tomza, and Gerritsma}}]{feldker2020buffer}
\bibinfo{author}{\bibfnamefont{T.}~\bibnamefont{Feldker}},
  \bibinfo{author}{\bibfnamefont{H.}~\bibnamefont{F{\"u}rst}},
  \bibinfo{author}{\bibfnamefont{H.}~\bibnamefont{Hirzler}},
  \bibinfo{author}{\bibfnamefont{N.}~\bibnamefont{Ewald}},
  \bibinfo{author}{\bibfnamefont{M.}~\bibnamefont{Mazzanti}},
  \bibinfo{author}{\bibfnamefont{D.}~\bibnamefont{Wiater}},
  \bibinfo{author}{\bibfnamefont{M.}~\bibnamefont{Tomza}}, \bibnamefont{and}
  \bibinfo{author}{\bibfnamefont{R.}~\bibnamefont{Gerritsma}},
  \bibinfo{journal}{Nat. Phys.} pp. \bibinfo{pages}{1--4}
  (\bibinfo{year}{2020}).

\bibitem[{\citenamefont{Ben-Shlomi et~al.}(2019)\citenamefont{Ben-Shlomi,
  Vexiau, Meir, Sikorsky, Akerman, Pinkas, Dulieu, and Ozeri}}]{ben2019strong}
\bibinfo{author}{\bibfnamefont{R.}~\bibnamefont{Ben-Shlomi}},
  \bibinfo{author}{\bibfnamefont{R.}~\bibnamefont{Vexiau}},
  \bibinfo{author}{\bibfnamefont{Z.}~\bibnamefont{Meir}},
  \bibinfo{author}{\bibfnamefont{T.}~\bibnamefont{Sikorsky}},
  \bibinfo{author}{\bibfnamefont{N.}~\bibnamefont{Akerman}},
  \bibinfo{author}{\bibfnamefont{M.}~\bibnamefont{Pinkas}},
  \bibinfo{author}{\bibfnamefont{O.}~\bibnamefont{Dulieu}}, \bibnamefont{and}
  \bibinfo{author}{\bibfnamefont{R.}~\bibnamefont{Ozeri}}
  (\bibinfo{year}{2019}), \eprint{1907.06736}.

\bibitem[{\citenamefont{H{\"a}rter and
  Hecker~Denschlag}(2014)}]{harter2014cold}
\bibinfo{author}{\bibfnamefont{A.}~\bibnamefont{H{\"a}rter}} \bibnamefont{and}
  \bibinfo{author}{\bibfnamefont{J.}~\bibnamefont{Hecker~Denschlag}},
  \bibinfo{journal}{Contemp. Phys.} \textbf{\bibinfo{volume}{55}},
  \bibinfo{pages}{33} (\bibinfo{year}{2014}).

\bibitem[{\citenamefont{Joger et~al.}(2017)\citenamefont{Joger, F{\"u}rst,
  Ewald, Feldker, Tomza, and Gerritsma}}]{joger2017observation}
\bibinfo{author}{\bibfnamefont{J.}~\bibnamefont{Joger}},
  \bibinfo{author}{\bibfnamefont{H.}~\bibnamefont{F{\"u}rst}},
  \bibinfo{author}{\bibfnamefont{N.}~\bibnamefont{Ewald}},
  \bibinfo{author}{\bibfnamefont{T.}~\bibnamefont{Feldker}},
  \bibinfo{author}{\bibfnamefont{M.}~\bibnamefont{Tomza}}, \bibnamefont{and}
  \bibinfo{author}{\bibfnamefont{R.}~\bibnamefont{Gerritsma}},
  \bibinfo{journal}{Phys. Rev. A} \textbf{\bibinfo{volume}{96}},
  \bibinfo{pages}{030703} (\bibinfo{year}{2017}).

\bibitem[{\citenamefont{C{\^o}t{\'e}}(2000)}]{cote2000classical}
\bibinfo{author}{\bibfnamefont{R.}~\bibnamefont{C{\^o}t{\'e}}},
  \bibinfo{journal}{Phys. Rev. Lett.} \textbf{\bibinfo{volume}{85}},
  \bibinfo{pages}{5316} (\bibinfo{year}{2000}).

\bibitem[{\citenamefont{C{\^o}t{\'e} and Dalgarno}(2000)}]{cote2000ultracold}
\bibinfo{author}{\bibfnamefont{R.}~\bibnamefont{C{\^o}t{\'e}}}
  \bibnamefont{and} \bibinfo{author}{\bibfnamefont{A.}~\bibnamefont{Dalgarno}},
  \bibinfo{journal}{Phys. Rev. A} \textbf{\bibinfo{volume}{62}},
  \bibinfo{pages}{012709} (\bibinfo{year}{2000}).

\bibitem[{\citenamefont{Lee et~al.}(2013)\citenamefont{Lee, Ravi, and
  Rangwala}}]{lee2013measurement}
\bibinfo{author}{\bibfnamefont{S.}~\bibnamefont{Lee}},
  \bibinfo{author}{\bibfnamefont{K.}~\bibnamefont{Ravi}}, \bibnamefont{and}
  \bibinfo{author}{\bibfnamefont{S.}~\bibnamefont{Rangwala}},
  \bibinfo{journal}{Phys. Rev. A} \textbf{\bibinfo{volume}{87}},
  \bibinfo{pages}{052701} (\bibinfo{year}{2013}).

\bibitem[{\citenamefont{Dutta and Rangwala}(2018)}]{dutta2018cooling}
\bibinfo{author}{\bibfnamefont{S.}~\bibnamefont{Dutta}} \bibnamefont{and}
  \bibinfo{author}{\bibfnamefont{S.}~\bibnamefont{Rangwala}},
  \bibinfo{journal}{Phys. Rev. A} \textbf{\bibinfo{volume}{97}},
  \bibinfo{pages}{041401} (\bibinfo{year}{2018}).

\bibitem[{\citenamefont{Zhang et~al.}(2009)\citenamefont{Zhang, Bodo, and
  Dalgarno}}]{zhang2009near}
\bibinfo{author}{\bibfnamefont{P.}~\bibnamefont{Zhang}},
  \bibinfo{author}{\bibfnamefont{E.}~\bibnamefont{Bodo}}, \bibnamefont{and}
  \bibinfo{author}{\bibfnamefont{A.}~\bibnamefont{Dalgarno}},
  \bibinfo{journal}{J. Phys. Chem. A} \textbf{\bibinfo{volume}{113}},
  \bibinfo{pages}{15085} (\bibinfo{year}{2009}).

\bibitem[{\citenamefont{Schmid et~al.}(2018)\citenamefont{Schmid, Veit, Zuber,
  L{\"o}w, Pfau, Tarana, and Tomza}}]{schmid2018rydberg}
\bibinfo{author}{\bibfnamefont{T.}~\bibnamefont{Schmid}},
  \bibinfo{author}{\bibfnamefont{C.}~\bibnamefont{Veit}},
  \bibinfo{author}{\bibfnamefont{N.}~\bibnamefont{Zuber}},
  \bibinfo{author}{\bibfnamefont{R.}~\bibnamefont{L{\"o}w}},
  \bibinfo{author}{\bibfnamefont{T.}~\bibnamefont{Pfau}},
  \bibinfo{author}{\bibfnamefont{M.}~\bibnamefont{Tarana}}, \bibnamefont{and}
  \bibinfo{author}{\bibfnamefont{M.}~\bibnamefont{Tomza}},
  \bibinfo{journal}{Phys. Rev. Lett.} \textbf{\bibinfo{volume}{120}},
  \bibinfo{pages}{153401} (\bibinfo{year}{2018}).

\bibitem[{\citenamefont{Schmidt-Mink et~al.}(1985)\citenamefont{Schmidt-Mink,
  M{\"u}ller, and Meyer}}]{schmidt1985ground}
\bibinfo{author}{\bibfnamefont{I.}~\bibnamefont{Schmidt-Mink}},
  \bibinfo{author}{\bibfnamefont{W.}~\bibnamefont{M{\"u}ller}},
  \bibnamefont{and} \bibinfo{author}{\bibfnamefont{W.}~\bibnamefont{Meyer}},
  \bibinfo{journal}{Chem. Phys.} \textbf{\bibinfo{volume}{92}},
  \bibinfo{pages}{263} (\bibinfo{year}{1985}).

\bibitem[{\citenamefont{Magnier et~al.}(1999)\citenamefont{Magnier, Rousseau,
  Allouche, Hadinger, and Aubert-Fr{\'e}con}}]{magnier1999potential}
\bibinfo{author}{\bibfnamefont{S.}~\bibnamefont{Magnier}},
  \bibinfo{author}{\bibfnamefont{S.}~\bibnamefont{Rousseau}},
  \bibinfo{author}{\bibfnamefont{A.}~\bibnamefont{Allouche}},
  \bibinfo{author}{\bibfnamefont{G.}~\bibnamefont{Hadinger}}, \bibnamefont{and}
  \bibinfo{author}{\bibfnamefont{M.}~\bibnamefont{Aubert-Fr{\'e}con}},
  \bibinfo{journal}{Chem. Phys.} \textbf{\bibinfo{volume}{246}},
  \bibinfo{pages}{57} (\bibinfo{year}{1999}).

\bibitem[{\citenamefont{Bouzouita et~al.}(2006)\citenamefont{Bouzouita, Ghanmi,
  and Berriche}}]{bouzouita2006ab}
\bibinfo{author}{\bibfnamefont{H.}~\bibnamefont{Bouzouita}},
  \bibinfo{author}{\bibfnamefont{C.}~\bibnamefont{Ghanmi}}, \bibnamefont{and}
  \bibinfo{author}{\bibfnamefont{H.}~\bibnamefont{Berriche}},
  \bibinfo{journal}{J. Mol. Struct. THEOCHEM} \textbf{\bibinfo{volume}{777}},
  \bibinfo{pages}{75} (\bibinfo{year}{2006}).

\bibitem[{\citenamefont{Jasik et~al.}(2007)\citenamefont{Jasik, Wilczy{\'n}ski,
  and Sienkiewicz}}]{jasik2007calculation}
\bibinfo{author}{\bibfnamefont{P.}~\bibnamefont{Jasik}},
  \bibinfo{author}{\bibfnamefont{J.}~\bibnamefont{Wilczy{\'n}ski}},
  \bibnamefont{and}
  \bibinfo{author}{\bibfnamefont{J.}~\bibnamefont{Sienkiewicz}},
  \bibinfo{journal}{Eur. Phys. J.-Spec.Top.} \textbf{\bibinfo{volume}{144}},
  \bibinfo{pages}{85} (\bibinfo{year}{2007}).

\bibitem[{\citenamefont{Musia{\l} et~al.}(2015)\citenamefont{Musia{\l}, Medrek,
  and Kucharski}}]{musial2015potential}
\bibinfo{author}{\bibfnamefont{M.}~\bibnamefont{Musia{\l}}},
  \bibinfo{author}{\bibfnamefont{M.}~\bibnamefont{Medrek}}, \bibnamefont{and}
  \bibinfo{author}{\bibfnamefont{S.~A.} \bibnamefont{Kucharski}},
  \bibinfo{journal}{Mol. Phys.} \textbf{\bibinfo{volume}{113}},
  \bibinfo{pages}{2943} (\bibinfo{year}{2015}).

\bibitem[{\citenamefont{Rabli and McCarroll}(2017)}]{rabli2017revised}
\bibinfo{author}{\bibfnamefont{D.}~\bibnamefont{Rabli}} \bibnamefont{and}
  \bibinfo{author}{\bibfnamefont{R.}~\bibnamefont{McCarroll}},
  \bibinfo{journal}{Chem. Phys.} \textbf{\bibinfo{volume}{487}},
  \bibinfo{pages}{23} (\bibinfo{year}{2017}).

\bibitem[{\citenamefont{Nasiri and Zahedi}(2017)}]{nasiri2017benchmark}
\bibinfo{author}{\bibfnamefont{S.}~\bibnamefont{Nasiri}} \bibnamefont{and}
  \bibinfo{author}{\bibfnamefont{M.}~\bibnamefont{Zahedi}},
  \bibinfo{journal}{Comput. Theor. Chem.} \textbf{\bibinfo{volume}{1114}},
  \bibinfo{pages}{106} (\bibinfo{year}{2017}).

\bibitem[{\citenamefont{Werner et~al.}(2012)\citenamefont{Werner, Knowles,
  Knizia, Manby, and Sch{\"u}tz}}]{MOLPRO-WIREs}
\bibinfo{author}{\bibfnamefont{H.-J.} \bibnamefont{Werner}},
  \bibinfo{author}{\bibfnamefont{P.~J.} \bibnamefont{Knowles}},
  \bibinfo{author}{\bibfnamefont{G.}~\bibnamefont{Knizia}},
  \bibinfo{author}{\bibfnamefont{F.~R.} \bibnamefont{Manby}}, \bibnamefont{and}
  \bibinfo{author}{\bibfnamefont{M.}~\bibnamefont{Sch{\"u}tz}},
  \bibinfo{journal}{WIREs Comput Mol Sci} \textbf{\bibinfo{volume}{2}},
  \bibinfo{pages}{242} (\bibinfo{year}{2012}).

\bibitem[{\citenamefont{Werner and Knowles}(1988)}]{werner1988efficient}
\bibinfo{author}{\bibfnamefont{H.-J.} \bibnamefont{Werner}} \bibnamefont{and}
  \bibinfo{author}{\bibfnamefont{P.~J.} \bibnamefont{Knowles}},
  \bibinfo{journal}{J. Chem. Phys.} \textbf{\bibinfo{volume}{89}},
  \bibinfo{pages}{5803} (\bibinfo{year}{1988}).

\bibitem[{\citenamefont{Prascher et~al.}(2011)\citenamefont{Prascher, Woon,
  Peterson, Dunning, and Wilson}}]{prascher2011gaussian}
\bibinfo{author}{\bibfnamefont{B.~P.} \bibnamefont{Prascher}},
  \bibinfo{author}{\bibfnamefont{D.~E.} \bibnamefont{Woon}},
  \bibinfo{author}{\bibfnamefont{K.~A.} \bibnamefont{Peterson}},
  \bibinfo{author}{\bibfnamefont{T.~H.} \bibnamefont{Dunning}},
  \bibnamefont{and} \bibinfo{author}{\bibfnamefont{A.~K.}
  \bibnamefont{Wilson}}, \bibinfo{journal}{Theoretical Chemistry Accounts}
  \textbf{\bibinfo{volume}{128}}, \bibinfo{pages}{69} (\bibinfo{year}{2011}).

\bibitem[{\citenamefont{Drake}(2006)}]{drake2006springer}
\bibinfo{author}{\bibfnamefont{G.~W.} \bibnamefont{Drake}},
  \emph{\bibinfo{title}{Springer handbook of atomic, molecular, and optical
  physics}} (\bibinfo{publisher}{Springer Science \& Business Media},
  \bibinfo{year}{2006}).

\bibitem[{\citenamefont{Yan and Drake}(1995)}]{yan1995eigenvalues}
\bibinfo{author}{\bibfnamefont{Z.-C.} \bibnamefont{Yan}} \bibnamefont{and}
  \bibinfo{author}{\bibfnamefont{G.~W.} \bibnamefont{Drake}},
  \bibinfo{journal}{Phys. Rev. A} \textbf{\bibinfo{volume}{52}},
  \bibinfo{pages}{3711} (\bibinfo{year}{1995}).

\bibitem[{\citenamefont{Aymar and Dulieu}(2005)}]{aymar2005calculation}
\bibinfo{author}{\bibfnamefont{M.}~\bibnamefont{Aymar}} \bibnamefont{and}
  \bibinfo{author}{\bibfnamefont{O.}~\bibnamefont{Dulieu}},
  \bibinfo{journal}{J. Chem. Phys.} \textbf{\bibinfo{volume}{122}},
  \bibinfo{pages}{204302} (\bibinfo{year}{2005}).

\bibitem[{\citenamefont{C{\^o}t{\'e}}(2016)}]{cote2016ultracold}
\bibinfo{author}{\bibfnamefont{R.}~\bibnamefont{C{\^o}t{\'e}}}, in
  \emph{\bibinfo{booktitle}{Advances In Atomic, Molecular, and Optical
  Physics}}, edited by
  \bibinfo{editor}{\bibfnamefont{E.}~\bibnamefont{Arimondo}},
  \bibinfo{editor}{\bibfnamefont{C.~C.} \bibnamefont{Lin}}, \bibnamefont{and}
  \bibinfo{editor}{\bibfnamefont{S.~F.} \bibnamefont{Yelin}}
  (\bibinfo{publisher}{Academic Press}, \bibinfo{year}{2016}),
  vol.~\bibinfo{volume}{65} of \emph{\bibinfo{series}{Advances In Atomic,
  Molecular, and Optical Physics}}, pp. \bibinfo{pages}{67--126}.

\bibitem[{\citenamefont{Tang et~al.}(2009)\citenamefont{Tang, Yan, Shi, and
  Babb}}]{tang2009nonrelativistic}
\bibinfo{author}{\bibfnamefont{L.-Y.} \bibnamefont{Tang}},
  \bibinfo{author}{\bibfnamefont{Z.-C.} \bibnamefont{Yan}},
  \bibinfo{author}{\bibfnamefont{T.-Y.} \bibnamefont{Shi}}, \bibnamefont{and}
  \bibinfo{author}{\bibfnamefont{J.~F.} \bibnamefont{Babb}},
  \bibinfo{journal}{Phys. Rev. A} \textbf{\bibinfo{volume}{79}},
  \bibinfo{pages}{062712} (\bibinfo{year}{2009}).

\bibitem[{\citenamefont{Mitroy et~al.}(2010)\citenamefont{Mitroy, Safronova,
  and Clark}}]{mitroy2010theory}
\bibinfo{author}{\bibfnamefont{J.}~\bibnamefont{Mitroy}},
  \bibinfo{author}{\bibfnamefont{M.~S.} \bibnamefont{Safronova}},
  \bibnamefont{and} \bibinfo{author}{\bibfnamefont{C.~W.} \bibnamefont{Clark}},
  \bibinfo{journal}{J. Phys. B} \textbf{\bibinfo{volume}{43}},
  \bibinfo{pages}{202001} (\bibinfo{year}{2010}).

\bibitem[{\citenamefont{Bardsley et~al.}(1975)\citenamefont{Bardsley, Holstein,
  Junker, and Sinha}}]{bardsley1975calculations}
\bibinfo{author}{\bibfnamefont{J.}~\bibnamefont{Bardsley}},
  \bibinfo{author}{\bibfnamefont{T.}~\bibnamefont{Holstein}},
  \bibinfo{author}{\bibfnamefont{B.}~\bibnamefont{Junker}}, \bibnamefont{and}
  \bibinfo{author}{\bibfnamefont{S.}~\bibnamefont{Sinha}},
  \bibinfo{journal}{Phys. Rev. A} \textbf{\bibinfo{volume}{11}},
  \bibinfo{pages}{1911} (\bibinfo{year}{1975}).

\bibitem[{\citenamefont{Bernheim et~al.}(1984)\citenamefont{Bernheim, Gold,
  Tipton, and Konowalow}}]{bernheim1984ionization}
\bibinfo{author}{\bibfnamefont{R.}~\bibnamefont{Bernheim}},
  \bibinfo{author}{\bibfnamefont{L.}~\bibnamefont{Gold}},
  \bibinfo{author}{\bibfnamefont{T.}~\bibnamefont{Tipton}}, \bibnamefont{and}
  \bibinfo{author}{\bibfnamefont{D.}~\bibnamefont{Konowalow}},
  \bibinfo{journal}{Chem. Phys. Lett.} \textbf{\bibinfo{volume}{105}},
  \bibinfo{pages}{201} (\bibinfo{year}{1984}).

\bibitem[{\citenamefont{Bernheim et~al.}(1983)\citenamefont{Bernheim, Gold, and
  Tipton}}]{bernheim1983rydberg}
\bibinfo{author}{\bibfnamefont{R.}~\bibnamefont{Bernheim}},
  \bibinfo{author}{\bibfnamefont{L.}~\bibnamefont{Gold}}, \bibnamefont{and}
  \bibinfo{author}{\bibfnamefont{T.}~\bibnamefont{Tipton}},
  \bibinfo{journal}{J. Chem. Phys.} \textbf{\bibinfo{volume}{78}},
  \bibinfo{pages}{3635} (\bibinfo{year}{1983}).

\bibitem[{\citenamefont{Le~Roy}(2017)}]{leroy2017level}
\bibinfo{author}{\bibfnamefont{R.~J.} \bibnamefont{Le~Roy}},
  \bibinfo{journal}{J. Quant. Spectrosc. Radiat. Transf.}
  \textbf{\bibinfo{volume}{186}}, \bibinfo{pages}{167 } (\bibinfo{year}{2017}).

\bibitem[{\citenamefont{Watson}(1980)}]{watson1980isotope}
\bibinfo{author}{\bibfnamefont{J.~K.} \bibnamefont{Watson}},
  \bibinfo{journal}{J. Mol. Spectrosc.} \textbf{\bibinfo{volume}{80}},
  \bibinfo{pages}{411} (\bibinfo{year}{1980}).

\bibitem[{\citenamefont{Joshi~et al.}(2020)}]{Nishant2019homonuclear}
\bibinfo{author}{\bibfnamefont{N.}~\bibnamefont{Joshi~et al.}},
  \bibinfo{journal}{unpublished}  (\bibinfo{year}{2020}).

\bibitem[{\citenamefont{Lutz and Hutson}(2016)}]{lutz2016deviations}
\bibinfo{author}{\bibfnamefont{J.~J.} \bibnamefont{Lutz}} \bibnamefont{and}
  \bibinfo{author}{\bibfnamefont{J.~M.} \bibnamefont{Hutson}},
  \bibinfo{journal}{J. Mol. Spectrosc.} \textbf{\bibinfo{volume}{330}},
  \bibinfo{pages}{43} (\bibinfo{year}{2016}).

\end{thebibliography}

\end{document}